\documentclass[twocolumn,prl,amsmath,amssymb]{revtex4-1}

\usepackage{amsthm}
\usepackage{graphicx}
\usepackage{color}
\usepackage{dcolumn}
\usepackage{bm}
\usepackage{braket}
\usepackage{float}
\usepackage[table]{xcolor}

\usepackage[utf8]{inputenc} 
\usepackage[T1]{fontenc}    
\usepackage{hyperref}       
\usepackage{url}            
\usepackage{booktabs}       
\usepackage{amsfonts}       
\usepackage{nicefrac}       
\usepackage{microtype}      
\usepackage{mathtools}      
\usepackage{mathrsfs}       
\usepackage{bbm} 
\usepackage{algorithm, algpseudocode}

\usepackage{hyperref}
\hypersetup{
    colorlinks = true,
    linkcolor = {blue},
    citecolor = {blue},
    urlcolor = {blue},
}

\newcommand{\sch}{Schr\"{o}dinger~}

\newcommand{\bb}{\ensuremath{\boldsymbol{b}}}
\newcommand{\bc}{\ensuremath{\boldsymbol{c}}}
\newcommand{\bw}{\ensuremath{\boldsymbol{w}}}
\newcommand{\bg}{\ensuremath{\boldsymbol{g}}}
\newcommand{\bp}{\ensuremath{\boldsymbol{p}}}
\newcommand{\bq}{\ensuremath{\boldsymbol{q}}}
\newcommand{\bh}{\ensuremath{\boldsymbol{h}}}

\newcommand{\by}{\ensuremath{\boldsymbol{y}}}
\newcommand{\bz}{\ensuremath{\boldsymbol{z}}}
\newcommand{\bo}{\ensuremath{\boldsymbol{o}}}
\newcommand{\bu}{\ensuremath{\boldsymbol{u}}}
\newcommand{\br}{\ensuremath{\boldsymbol{r}}}

\newcommand{\hbo}{\ensuremath{\hat{\boldsymbol{o}}}}

\newcommand{\bI}{\ensuremath{\boldsymbol{I}}}
\newcommand{\bX}{\ensuremath{\boldsymbol{X}}}

\newcommand{\bF}{\ensuremath{\boldsymbol{F}}}
\newcommand{\bP}{\ensuremath{\boldsymbol{P}}}
\newcommand{\bW}{\ensuremath{\boldsymbol{W}}}

\newcommand{\cM}{\ensuremath{\mathcal{M}}}
\newcommand{\cR}{\ensuremath{\mathcal{R}}}
\newcommand{\cH}{\ensuremath{\mathcal{H}}}
\newcommand{\cQ}{\ensuremath{\mathcal{Q}}}
\newcommand{\cT}{\ensuremath{\mathcal{T}}}
\newcommand{\cL}{\ensuremath{\mathcal{L}}}
\newcommand{\cN}{\ensuremath{\mathcal{N}}}

\newcommand{\sQ}{\ensuremath{\mathscr{Q}}}
\newcommand{\hsig}{\ensuremath{\sigma}}

\newcommand{\ehp}{\ensuremath{e^{iH\tau}}}
\newcommand{\ehm}{\ensuremath{e^{-iH\tau}}}
\newcommand{\ehmt}{\ensuremath{e^{-iH{(\tau/V)}}}}
\newcommand{\wfeed}{\ensuremath{\bW_{\textup{con}}}}
\newcommand{\wout}{\ensuremath{\bW_{\textup{out}}}}

\newcommand{\nqrc}{\ensuremath{N_{\textup{qr}}}}
\newcommand{\nin}{\ensuremath{N_{\textup{in}}}}
\newcommand{\nout}{\ensuremath{N_{\textup{out}}}}
\newcommand{\noutv}{\ensuremath{N_{\textup{out}}V}}
\newcommand{\ntot}{\ensuremath{N_{\textup{total}}}}
\newcommand{\mc}{\textup{MC}}
\newcommand{\mf}{\textup{MF}}

\DeclareMathOperator{\tr}{\textup{Tr}}
\DeclareMathOperator{\nrmse}{\textup{NRMSE}}
\DeclareMathOperator{\vpt}{\textup{VPT}}
\DeclareMathOperator{\vc}{\textup{vec}}

\newtheorem{definition}{Definition}

\begin{document}
\title{Higher-order quantum reservoir computing}

\author{Quoc Hoan Tran}
\email{tran\_qh@ai.u-tokyo.ac.jp}

\author{Kohei Nakajima}
\email{k\_nakajima@mech.t.u-tokyo.ac.jp}
\affiliation{
	Graduate School of Information Science and Technology, The University of Tokyo, Tokyo 113-8656, Japan
}

\date{\today}

\begin{abstract}
Quantum reservoir computing (QRC) is an emerging paradigm for harnessing the natural dynamics of quantum systems as computational resources that can be used for temporal machine learning tasks.
In the current setup, QRC is difficult to deal with high-dimensional data and has a major drawback of scalability in physical implementations. 
We propose higher-order QRC, a hybrid quantum-classical framework consisting of multiple but small quantum systems that are mutually communicated via classical connections like linear feedback.
By utilizing the advantages of both classical and quantum techniques, our framework enables an efficient implementation to boost the scalability and performance of QRC.
Furthermore, higher-order settings allow us to implement a FORCE learning or an innate training scheme, which provides flexibility and high operability to harness high-dimensional quantum dynamics and significantly extends the application domain of QRC.
We demonstrate the effectiveness of our framework in emulating large-scale nonlinear dynamical systems, including complex spatiotemporal chaos, which outperforms many of the existing machine learning techniques in certain situations.
\end{abstract}

\maketitle

\section{Introduction}
It is postulated that quantum computers may outperform classical computers when it comes to machine learning (ML) tasks due to the superior ability of quantum mechanics 
to generate counter-intuitive patterns~\cite{biamonte:2017:QML}.
Quantum machine learning (QML) is an active interdisciplinary research area proposed from this motivation to improve existing ML methods through the advantages of quantum mechanics~\cite{ciliberto:2018:QML}.
For the foreseeable future, QML algorithms are expected to run on noisy intermediate-scale quantum (NISQ) devices, which includes a few tens of qubits and supports only non-error-corrected computations~\cite{preskill:2018:NISQ}.
Existing quantum techniques are utilized to their fullest extent in ML tasks on these devices via hybrid quantum–classical methods that combine classical learning regimes with the advantages of quantum systems~\cite{dassarma:2019:MLQPhys,torlai:2020:NISQ}.

\begin{figure*}\label{fig:hqrc}
  \centering
  \includegraphics[width=0.9\linewidth]{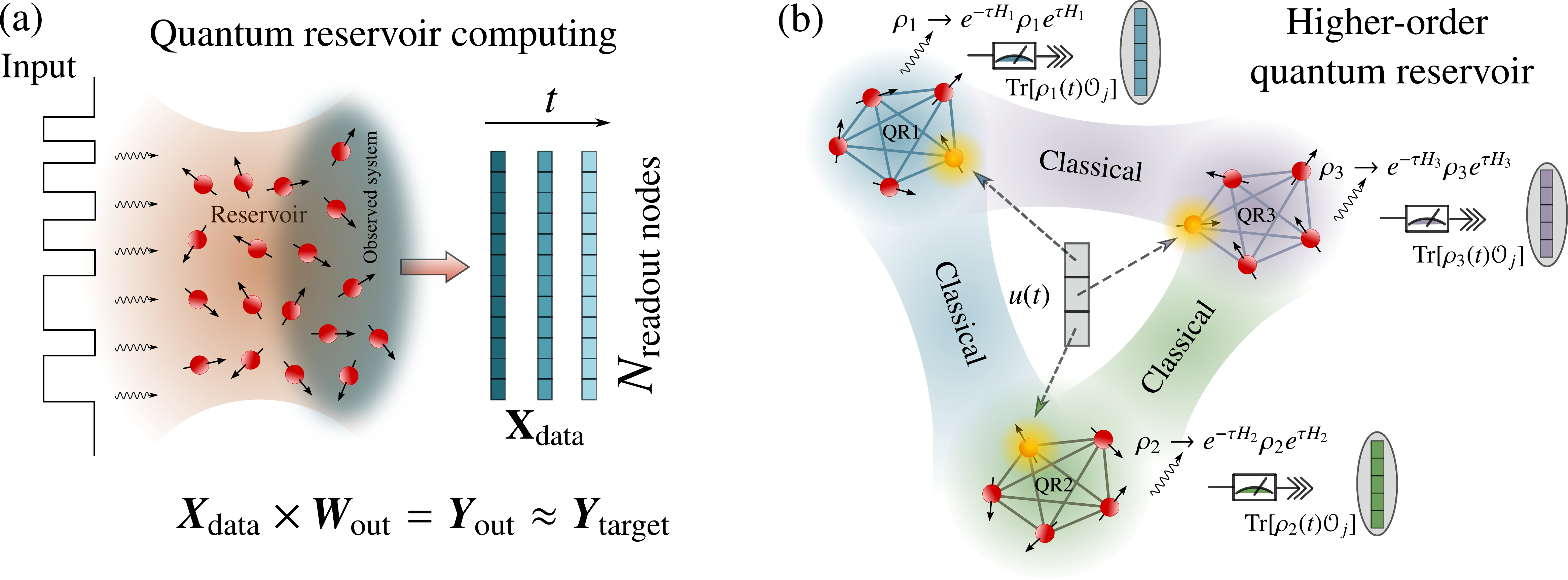}
  \caption{Schematic of quantum reservoir computing (QRC) and the higher-order quantum reservoir (HQR). (a) The input stream drives the transition of the state of qubits, and the quantum system continues evolving itself for a specific time interval to spread the information of the input. Measurements are performed on some qubits, which correspond to signals of readout nodes. After collecting a sufficient amount of input-output signal pairs, the readout weights are trained to emulate the target signals. (b) The HQR consists of multiple different quantum reservoirs that are mutually connected via classical connections to enable multidimensional inputs and enhance both quantum and classical advantages.}
\end{figure*}

Reservoir computing (RC)~\cite{maass:2002:rc,jaeger:2004:rc,verstraeten:2007:esn,lukosevicius:2009:rcrecurr} is a framework that originated from recurrent neural networks to efficiently solve temporal ML tasks.
Conventional RC consists of a randomly connected network called a reservoir and a trainable readout part for pattern analysis from output states of the reservoir.
The input stream is fed into the reservoir, which functions like a nonlinear processing unit to project low-dimensional input into a high-dimensional dynamical system.
Since the training in RC is simple and extremely fast, it is highly suitable and amendable for hardware implementation in a wide variety of physical systems~\cite{appeltant:2011:information,larger:2012:optic,torrejon:2017:neuromorphic,furuta:2018:spin,nakajima:2013:soft,nakajima:2018:exploiting,nakajima:2020:physrc}.
Quantum reservoir computing (QRC) is a variation of RC, where the reservoir is implemented as a quantum many-body system such as a set of interacting qubits~\cite{fujii:2017:qrc,nakajima:2019:qrc} or a set of fermions~\cite{ghosh:2019:quantum,ghosh:2019:neuromorphic,ghosh:2020:reconstruct} driven by a Hamiltonian dynamics.
The random connections in the classical reservoir are replaced by basic quantum tunneling.
The input stream then drives the transition state, which is evolved through a unitary operator based on the dynamics of the system.
Thereafter, quantum measurements are performed to obtain signals that can be considered as reservoir states for training [Fig.~\ref{fig:hqrc}(a)].
In this respect, QRC does not require a quantum computer's pre-existence but uses the analog quantum mechanical system for temporal processing tasks.

The nuclear magnetic resonance (NMR) spin ensemble in a molecular solid 
has been reported as the first physical implementation of QRC~\cite{negoro:2018:spins}.
In this implementation, each QRC is an ensemble quantum system consisting of many copies of the same molecule.
Signals are averaged from the massive number of copies of molecules in the system without concern for the effects of projective measurements.
In the NMR system, the local control and measurement of a qubit are performed using the difference in the nuclear species' resonant frequency.
However, only a few species should be chosen for use in an NMR system. 
It is not easy to design and synthesize a molecule that includes many addressable spins with diﬀerent species and in different environments.
For example, the current limit is a 12-addressable-spin system in a liquid~\cite{negrever:2006:12qbit,lu:2017:npj:12qbit}.
Therefore, increasing the scalability of the NMR-based implementation requires the enlargement and redesign of sample molecules, which is not feasible in practice due to the operational cost.
Moreover, a multidimensional input via a single QRC requires perfect synchronization between the input-induced spins, which is operationally difficult to achieve in a single system.

In this paper, we propose a general hybrid quantum-classical framework for RC, namely, higher-order quantum reservoir computing (HQRC).
We aim to utilize ensemble small quantum systems as a big reservoir and enhance its computational power through random connections between sub-systems [Fig.~\ref{fig:hqrc}(b)].
Due to the operational difference between preparing one huge quantum system and preparing multiple small-size quantum systems, 
our framework enables a realistic and practical physical implementation of QRC.
This scheme is called ``higher-order'' in terms of different quantum systems placed in an ensemble reservoir, where each system can be considered as a node in the reservoir.
By introducing the higher-order system where each system receives a single input component, we can coordinate I/O  timesteps among systems, and then the multidimensional inputs could be easily applied.
The input of each sub-system is a linear combination of the common input streams and the signals obtained from other systems.  This scheme enables to equip a massive amount of computational nodes with the controllable linear feedback, which is expected to increase the expressive power of the system.
Furthermore, higher-order settings enable the implementation of FORCE learning~\cite{sussillo:2009:FORCE} and innate training~\cite{laje:2013:innate} in the quantum context to broaden the applications of QRC.

In the following sections, we provide a detailed procedure for the temporal processing in HQRC with theoretical explanations of parameter design, which are useful for evaluating a quantum reservoir (QR) as a learning system.
We also provide fundamental but novel theoretical results in understanding the dynamics of QRC, then numerically verify that the computational power of QRC, like the memory capacity, can be enhanced through higher-order settings.
We further present the implementation of quantum-innate training based on FORCE learning to enhance the noise robustness of the reservoir signals in higher-order settings.
Finally, we demonstrate the effectiveness of our proposal in emulating nonlinear dynamical systems, including high-dimensional spatiotemporal chaos.
Interestingly, the experimental results indicate that our approach performs comparatively with classical approaches and even outperforms them in certain situations, such as in the limitation of the training data.
From the physical implementation perspective, our framework also paves several opportunities for effective design of recently proposed experimental platforms for QRC~\cite{negoro:2018:spins,chen:2020:temporal}.

\section{Preliminaries}
We describe some standard notions that are necessary in our analysis~\cite{nielsen:2011:QCQI}.
First, we define $\cM_{m\times n}$ and $\cR_{m\times n}$ as the sets of $m\times n$-dimensional matrix with complex elements and real elements, respectively.
For $U \in \cM_{m\times n}$, we denote $U^\dagger \in \cM_{n\times m}$ as the conjugate transpose of $U$.
A complex square matrix $U$ is called a unitary matrix if $UU^\dagger= U^\dagger U$ is the identity matrix, and a Hermitian matrix if $U=U^\dagger$.
The tensor product of two matrices $A \in \cM_{p\times q}$ and $B \in \cM_{r\times s}$ is defined as the $pr \times qs$ block matrix $A\otimes B = [a_{ij}B]_{i,j=1}^{p,q}$.

\paragraph{Pure and mixed states. }
A pure state $\ket{\psi}$ of a spin is a two-dimensional complex vector spanned by the eigenstates $\{\ket{0}, \ket{1}\}$ of the Pauli operator $\hsig^z = \left[\begin{array}{cc}1 & 0\\0 & -1\end{array}\right]$.
We denote $\bra{\psi}=(\ket{\psi})^\dagger$; therefore,
$\ket{0}=\left[\begin{array}{c} 1\\0 \end{array}\right], \bra{0}=\left[1 \quad 0\right]$, and $\ket{1}=\left[\begin{array}{c} 0\\1 \end{array}\right], \bra{1}=\left[0 \quad 1\right]$.
Given a quantum system $S$ comprising $N$ qubits, the Hilbert space $\cH_S$ of $S$ is a tensor product space of two-dimensional individual spin spaces.
A pure quantum state is represented by a normalized $2^N$-dimensional vector $\ket{\Psi}$. 
Generally, the quantum state can be considered a statistical mixture of pure states, which is described by a density matrix $\rho \in \cM_{2^N\times2^N}$. 
For a pure state $\ket{\Psi}$, the density matrix is defined as $\rho=\ket{\Psi}\bra{\Psi}$.
If we consider a quantum state of the system that is sampled from a set of pure states $\{\ket{\Psi_k}\}$ with a probability distribution $\{p_k\}$, the density matrix is given by $\rho=\sum_kp_k\ket{\Psi_k}\bra{\Psi_k}$.
The density matrix $\rho$ is Hermitian and positive semidefinite; moreover, $\tr(\rho)=1$, where $\tr(X)$ denotes the trace of square matrix $X$.

\paragraph{Partial trace. }
Given a composite system $S$ of two subsystems $S_1$ and $S_2$, the Hilbert space of $S$ is the tensor product $\cH_{S_1}\otimes\cH_{S_2}$.
From the density matrix $\rho$ of $S$, we can recover the marginal ``reduced density matrix'' for a subsystem via the partial trace operation $\tr_1(\rho)$,
which is defined as the linear extension of the mapping $\tr_1: A_1 \otimes A_2 \mapsto \tr(A_1)A_2$ for matrices $A_1\in\cH_{S_1}, A_2\in\cH_{S_2}$.

\paragraph{Observables and time evolution. }
Observables of quantum system $S$ correspond to Hermitian matrices $O \in \cM_{2^N\times2^N}$.
If we perform measurements on these observables at state $\rho$,
it means that $S$ is interacting with measuring apparatus in the presence of the surrounding environment.
The outcomes of the measurement are recorded on the dial on the measuring apparatus, and 
the statistics of measurement outcomes is determined by the expected value of $O$ as $\langle O \rangle = \tr[\rho O]$.
If $S$ is a closed system, the time evolution is generated by a Hamiltonian $H$ via the \sch equation
$
    i\dfrac{d}{dt}\ket{\Psi(t)} = H\ket{\Psi(t)},
$
where $H \in \cM_{2^N\times2^N}$ is the Hermitian matrix that defines the system dynamics.
For a time $\tau$, the evolution from $t$ to $t+\tau$ is given by $\ket{\Psi(t+\tau)} = U_{\tau}\ket{\Psi(t)}$, or by an expression in terms of density matrix $\rho(t+\tau) = U_{\tau}\rho(t)U_{\tau}^{\dagger}$, 
where $U_{\tau}=\ehm$ is a unitary matrix.
Most generally, the time evolution law is described by a map called a completely positive and trace preserving (CPTP) 
map $\cL: \rho \to \rho^{\prime}$ with the following properties: linear, trace preserving, Hermiticity preserving, and completely positive. Here, a map $\cL_1$ is completely positive if $\cL_1\otimes \mathbbm{1}_2$ is positive for any extension $\cH_2$ of the Hilbert space $\cH_1$.

\section{Quantum reservoir computing}
\subsection{Definition of Quantum-reservoir Dynamics}
We briefly explain the background of the quantum-reservoir dynamics used in QRC.
The extensive explanation of QRC can be found in Ref.~\cite{fujii:2017:qrc}. 
For simplicity, 
we consider a one-dimensional input and output case with the input sequence $u=\{u_1,\ldots,u_L\}$ and the corresponding target sequence $\hat{y}=\{\hat{y}_1,\ldots,\hat{y}_L\}$, 
where $u_k$ is a continuous variable in $[0, 1]$.
QRC emulates a nonlinear function $\sQ$ to produce the output $y_k=\sQ(\bw, \rho^{(0)}, \{u_l\}_{l=1}^k)$ ($k=1,\ldots,L$).
Here, $\rho^{(0)}$ is the initial state of the quantum system, and $\bw$ is the parameter that needs to be optimized.
A temporal learning task consists of three phases: a washout phase, a training phase, and an evaluation phase.
In the washout phase, the system evolves for the first $T$ transient  steps to washout the initial conditions from the dynamics.
The training phase to optimize $\bw$ is performed with training data $(\{u_k\}_{k=T}^{L_1}, \{y_k\}_{k=T}^{L_1})$, where $1\leq T < L_1 < L$, such that the mean-square error between $y_k$ and $\hat{y}_k$ over $k=T,\ldots,L_1$ becomes minimum.
The trained parameter $\bw$ is used to generate outputs in the evaluation phase.

For an $N$-qubits system, at time $t=(k-1)\tau$, the input $u_k\in [0, 1]$ is fed to the system by setting the density matrix of the first spin to 
\begin{align}\label{eqn:input:spin}
\varrho_{u_k}=(1-u_k)\ket{0}\bra{0} + u_k\ket{1}\bra{1} \in \cM_{2\times 2}.
\end{align}
Therefore, the density matrix $\rho \in \cM_{2^N\times 2^N}$ of the entire system is mapped by a CPTP map 
\begin{align}\label{eqn:input:tr}
    \rho \to \cT_{u_k}(\rho) = \varrho_{u_k} \otimes \tr_1[\rho],
\end{align}
where $\tr_1$ denotes a partial trace with respect to the first qubit.
After the input is set, the system continues evolving itself during time interval $\tau$. 
The dynamics are governed by the \sch equation and the information of the input sequence encoded in the first spin spreads through the system. 
It follows that the state of the system before the next input $u_{k+1}$ is
\begin{align}\label{eqn:input:exp}
    \rho^{(k)} = \ehm \cT_{u_k}(\rho^{(k-1)}) \ehp,
\end{align}
where $\rho^{(k)} = \rho(k\tau)$ is the density matrix at $t=k\tau$.
It is noted that the size of $\rho_k$ grows exponentially with the number of qubits.
We obtain partial information regarding $\rho^{(k)}$ by measuring local observables on qubits.
The signals for training are obtained from average values of measurement results on each qubit.
More precisely, if we employ the ordered basis $\{O_j\}$ in the operator space, 
then the observed signals at time $t$ are the first $N_{\textup{out}}$ elements $s_j(t)=\tr[\rho(t)O_j]$,
where the selection of observables depends on the physical implementation of the system.
For example, if we consider the spin receiving the input is an ancilla qubit that we cannot perform the measurement on it, the observed operators act only on the other spins.
Since this does not change our main points, 
for the sake of convenience, we consider a situation in which we can perform measurements on all qubits.

\subsection{Temporal Multiplexing}
In practical applications, the temporal multiplexing scheme is introduced to improve the performance in extracting dynamics.
Here, the signals are measured not only at time $k\tau$ but also at each of the subdivided $V$ time intervals during the evolution in interval $\tau$ to construct $V$ virtual nodes.
The density matrix is then updated by
\begin{align}
    \rho((k-1)\tau+\frac{1}{V}\tau) &= U_{(\tau/V)}\cT_{u_k}(\rho^{(k-1)})U_{(\tau/V)}^{\dagger}, \label{eqn:time:1}\\
	\rho((k-1)\tau+\frac{v}{V}\tau) &= \label{eqn:time:2}\\ U_{(\tau/V)}\rho(&(k-1)\tau+\frac{v-1}{V}\tau)U_{(\tau/V)}^{\dagger} (v=2,\ldots,V),\nonumber
\end{align}
where $U_{(\tau/V)}=\ehmt$.
Therefore, we can obtain temporal signals from $N_{\textup{out}}V$ nodes.
The learning procedure is straightforward as we 
parameterize the linear readout function as $y_k=\sum_{i=0}^{N_{\textup{out}}V}w_ix_{ki}$, where $x_{ki} = s_j((k-1)\tau + \frac{v}{V}\tau)$ for $i = (j-1)V + v > 0$ $(1\leq v \leq V, 1\leq j \leq N_{\textup{out}})$. Here, $x_{k0}=1.0$ are introduced as constant bias terms, and $\bw = [w_0 \quad w_1 \quad \ldots \quad w_{N_{\textup{out}}V}]$ represents the readout weight parameters.
If we denote $K$ as the number of time steps used in the training phase, $\bw$ is optimized via the linear regression, or the Ridge regression in the matrix form $\boldsymbol{\hat{w}^\top} = (\bX^\top\bX + \beta\bI)^{-1}\bX^\top\hat{\by}$.
Here, $\hat{\by}=[\hat{y}_1\quad\ldots\quad \hat{y}_K]^\top$ is the target sequence,
$\bX = (x_{ki})\in \cR_{K\times(\noutv + 1)}$ is the training data matrix 
and $\beta$ is the parameter
serves as the positive constant shifting the diagonals 
introduced to avoid the problem of the near-singular moment matrix.

\subsection{Model}
It is important to note that QR does not train the parameters of quantum systems except for the linear readout weights.
Therefore, any quantum system can be employed as long as it can exhibit sufficiently rich dynamics to fulfill the computational requirements for reservoir framework. 
We employ the fully connected transverse field Ising model,
which is the standard workhorse to build the QR.
The Hamiltonian is given by $
    H = J\sum_{i\neq j}h_{i,j}\hsig^x_i\hsig^x_j + J\sum_jg_j\hsig^z_j
$, where all the spins interact with each other in $x$-direction and are coupled to an external magnetic field in $z$-direction.
Here, $\hsig_j^\gamma$ $(\gamma \in \{x, y, z\})$ is the operator measuring the spin $j$ along the $\gamma$ direction, which can be described as an $N$-tensor product of $2\times 2$-matrices as
\begin{align}
\hsig_j^\gamma = \bI\otimes\ldots\otimes\underbrace{\hsig^{\gamma}}_\text{$j$-index}\otimes\ldots\otimes\bI,
\end{align}
where
$
\bI = \left[\begin{array}{cc}1 & 0\\0 & 1\end{array}\right]$,
$
\hsig^x = \left[\begin{array}{cc}0 & 1\\1 & 0\end{array}\right]$,
$
\hsig^y = \left[\begin{array}{cc}0 & -i\\i & 0\end{array}\right]$,
and
$
\hsig^z = \left[\begin{array}{cc}1 & 0\\0 & -1\end{array}\right].
$
$J$ is the coupling magnitude of the Hamiltonian,
while the coupling parameter $h_{i,j}$ and the transverse field parameter $g_j$ are taken uniformly from $[-1.0, 1.0]$.
We select $\nout=N$ observables $O_j=\sigma^z_j$ to produce the signals of readout nodes.
Since the maps in Eq.~\eqref{eqn:input:tr} and Eq.~\eqref{eqn:input:exp} are linear, it may appear that the system does not correspond to nonlinear dynamics. 
However, the nonlinearity is emergent from higher-order correlations or nonlinear terms, which are mixed by the linear but nonintegrable dynamics via $U_{\tau}$~\cite{fujii:2017:qrc}.
	
\section{Higher-order quantum reservoir computing}
\subsection{Higher-order Model}
We propose an effective and practical design to enhance the computational ability of the QR.
Our higher-order quantum reservoir (HQR) consists of an ensemble of $\nqrc$ QRs, such that the $l$th system $\cQ_l$ has the Hamiltonian $H_l$ with $N_l$ qubits (which are known as true nodes) and $V_l$ virtual nodes.
We denote HQR-$n$ as the HQR comprising $n$ QRs.
For one-dimensional input, these QR systems are driven by a common input stream, while they can receive different input streams for the multidimensional input setting in general.
Algorithm~\ref{alg:hqrc} presents the temporal processing steps of the HQR. 
Here, we denote $\rho^{(k)}_l$ as the density matrix of $\cQ_l$ at time $t=k\tau$, where $\rho^{(0)}_l$ is the initialized density matrix.
To fed a $\nin$-dimensional external input $\bu$ into the HQR, we can transform $\bu$ into a $\nqrc$-dimensional input via the linear transformation $\bu\to\bW_{\textup{in}}\bu$ where $\bW_{\textup{in}} \in \cR_{\nqrc \times \nin}$ is fixed randomly.
The reservoir states of $\cQ_l$ at time $t=k\tau$ ($k\geq 0$) are represented by a $V_lN_l$-dimensional vector $\bz_{kl}$, which is initialized at $k=0$ as the zero vector $\bz_0$.
Then the reservoir states of the HQR at $t=k\tau$ can be represented by a $\ntot$-dimensional vector $\bz_k=[\bz_{k1}^{\top},\ldots,\bz_{k\nqrc}^{\top}]^\top$, where $\ntot=\sum_{l=1}^{\nqrc} N_lV_l$.
We employ the scaling $z\to (z+1)/2$ such that the elements of $\bz$ are in $[0, 1]$.

At $t=(k-1)\tau$ ($k\geq 1$), the mixed input $u^{\prime}_{kl}$, which is injected into $Q_{l}$, is a linear combination of the external input and the reservoir states from all QRs. 
The mixed input $\bu^{\prime}_k \in \cR_{\nqrc\times 1}$ can be represented in the matrix form
\begin{align}
    \bu^{\prime}_k = (1-\alpha)\bW_{\textup{in}}\bu_k + \alpha\wfeed\bz_{k-1},
\end{align}
where $\wfeed \in \cR_{\nqrc\times\ntot}$ is randomly generated and fixed to represent the linear feedback connection between the QRs,
and $\alpha$ ($0\leq \alpha \leq 1$) is defined as the connection strength parameter.
Because our input-injection scheme in Eq.~\eqref{eqn:input:spin} requires the input in the interval $[0, 1]$, we must ensure that the elements of $\bu^{\prime}_k$ are in $[0, 1]$. 
Since the elements of $\bu_k$ and $\bz_k$ are in $[0, 1]$, we generate $\bW_{\textup{in}}$
and $\wfeed$ such that their elements are nonnegative and the total value of elements in each row is less than or equal to $1$.
In our experiments, 
we set up the matrix $\wfeed$ such that there is no self-loop connection in each QR
to compare the differences between the single QRC and the higher-order QRC. 
After injecting the mixed input $u^{\prime}_{kl}$ into $\cQ_l$, the density matrix in $\cQ_l$ is transformed by the CPTP map, $\cT_{u^{\prime}_{kl}}$, as in Eq.~\eqref{eqn:input:tr},
and is consequently evolved in each $\tau/V$ time, as in Eqs.~\eqref{eqn:time:1}\eqref{eqn:time:2}.
The training is performed with 
the temporal reservoir states of the entire system, which are rewritten with bias terms as the matrix $\bX \in \cR_{K\times(\ntot + 1)}$.

\begin{algorithm}[H]
\caption{Temporal processing of HQR}
\begin{algorithmic}[1]
\Require The input stream $\{\bu_k\}$ ($\bu_k\in\cR_{\nin\times1}$), $\nqrc$ QRs $\cQ_l$ with corresponding configurations ($H_l, N_l, V_l, \rho_l^{(0)}$), the time interval $\tau$, the connection strength $\alpha\in [0,1]$, the matrices $\bW_{\textup{in}}\in \cR_{\nqrc \times \nin}$ and $\wfeed\in \cR_{\nqrc\times\ntot}$.
The reservoir states $\bz_{kl} \in \cR_{(V_lN_l)\times 1}$ of $\cQ_l$, and the entire states $\bz_k=[\bz_{k1}^{\top},\ldots,\bz_{k\nqrc}^{\top}]^\top \in \cR_{\ntot\times 1}$ at $t=k\tau$. 
\For{input $\bu_k$ ($k\geq 1$)}
\State $\bu^{\prime}_k \leftarrow (1-\alpha)\bW_{\textup{in}}\bu_k + \alpha \wfeed\bz_{k-1}$
\For{$l=1,\ldots,\nqrc$}
\State $\rho_l \leftarrow \cT_{u^{\prime}_{kl}}(\rho_l^{(k-1)})$ \Comment{$\cT_{u^{\prime}_{kl}}$ is a CPTP map defined in Eq.~\eqref{eqn:input:tr}}
\For{$v=1,\ldots,V_l$}
\State $\rho_l \leftarrow  e^{-\frac{\tau}{V_l}H_l}\rho_le^{\frac{\tau}{V_l}H_l}$ \Comment{$\cQ_l$ is evolved in time interval $\tau/V_l$ }
\For{$j=1,\ldots,N_l$}
\State $z_{klvj}\leftarrow \tr[\rho_l\sigma_j^z]$ \Comment{Measure the average spin values in the $z$-direction}
\State $z_{klvj}\leftarrow (1+z_{klvj})/2$\Comment{Scaling}
\EndFor
\EndFor
\State $\bz_{kl} \leftarrow [(z_{klvj})_{vj}]^\top \in \cR_{( V_l N_l)\times 1}$ 
\State $\rho_l^{(k)} \leftarrow \rho_l$
\EndFor 
\State $\bz_k \leftarrow [(\bz_{k1})^\top \ldots (\bz_{k\nqrc})^\top]^\top \in \cR_{(\sum_{l=1}^{\nqrc} V_l N_l)\times 1}$ \Comment{The reservoir states at $t=k\tau$}
\EndFor
\State \Return Reservoir temporal states: $\bz = \{\bz_1, \bz_2, \ldots\}$
\end{algorithmic}\label{alg:hqrc}
\end{algorithm}

\subsection{Properties of HQR Dynamics}
In the classical regime, the reservoir is required to satisfy an asymptotic stability property termed the echo state property (ESP), which ensures that the computations are performed independently with the initial state of the reservoir~\cite{jaeger:2001:echo}.
ESP is a similar concept with the fading memory property, which states that the reservoir must produce the close outputs if the corresponding inputs are close in recent times~\cite{boyd:1985:fading}.
We explore the asymptotic behavior and memory capacity of HQR systems.

\paragraph{Asymptotic stability. }
A ``robust'' HQR must produce trajectories that are robust to small perturbations to the system--that is, the computations performed by an HQR system for the same input are independent of its initial density matrix. We define this as quantum echo state property (QESP), which is also mentioned in the convergence property of dissipative quantum systems in Ref.~\cite{chen:2019:dissipative}.

\begin{definition}
An HQR system whose reservoir dynamics are governed by Algorithm~\ref{alg:hqrc} is said to satisfy the quantum echo state property (QESP) with respect to the Schatten $p$-norm if for each initial density matrix $\rho^{(0)}, \hat{\rho}^{(0)}$, and for any input sequence $\bu_{L}=\{u_l\}_{l=1}^{L}$ of length $L$, it holds that $\Vert \rho^{(L)} - \hat{\rho}^{(L)} \Vert_p \to 0$ as $L\to \infty$.
Here, $\Vert \cdot \Vert_p$ denotes the Schatten $p$-norm for $p \geq 1$, defined as $\Vert A \Vert_p = \tr[(\sqrt{A^\dagger A})^p]^{1/p}$ for matrix $A$.
\end{definition}

\begin{algorithm}[H]
\caption{The QESP index}
\begin{algorithmic}[1]
\Require $\nqrc$ QR systems $\cQ_l$ ($l=1,\ldots,\nqrc$), number of washout time steps $T$, number of evaluation time steps $E$, number of trials $P$. Denote $\bF(\rho^{(0)}, \{\bu_i\}_{i=1}^k)\in\cR_{\ntot\times 1}$ as the reservoir states of the system at time $t=k\tau$ for input sequence $\{\bu_i\}_{i=1}^k$ and the initial density matrix $\rho^{(0)}$.
\State Initialize the density matrices $\rho^{(0)}_l$ ($l=1,\ldots,\nqrc)$.
\State $\rho^{(0)}\leftarrow \rho^{(0)}_1\otimes \ldots \rho^{(0)}_{\nqrc}$
\For{$p=1,\ldots,P$}
\State Initialize the density matrices $\sigma^{(0)}_l$ ($l=1,\ldots,\nqrc)$.
\State $\sigma^{(0)}\leftarrow \sigma^{(0)}_1\otimes \ldots \sigma^{(0)}_{\nqrc}$
\For{$k=1,\ldots,E$}
\State $\delta_p(k) = \Vert \bF(\rho^{(0)}, \{\bu_i\}_{i=1}^{k+T}) - \bF(\sigma^{(0)}, \{\bu_i\}_{i=1}^{k+T})\Vert $
\EndFor
\State $\mu_p = \langle \delta_p(k) \rangle_k$ \Comment{Average over the evaluations}
\EndFor
\State $\mu \leftarrow \langle \mu_p \rangle_p$
\Comment{Average over the trials}
\State \Return The QESP index $\mu$
\end{algorithmic}\label{alg:esp}
\end{algorithm}

If QESP is satisfied, it means that for the same input sequence, density operators asymptotically converge without depending on initial values. 
Here, we consider an empirical perspective to the study of the asymptotic stability.
We extend the algorithm in~\cite{gallicchio:2018:chasing} to define the QESP index to evaluate the average deviation of observed signals generated from random initial states to reference signals starting from a fixed state (see Algorithm~\ref{alg:esp} for details).
The first $T$ time steps are discarded as the washout phase,
and the QESP index is averaged over $P$ randomly generated initial density matrices for the range of dynamics in $E$ evaluation time steps.

\begin{figure}
	\centering
	\includegraphics[width=1.0\linewidth]{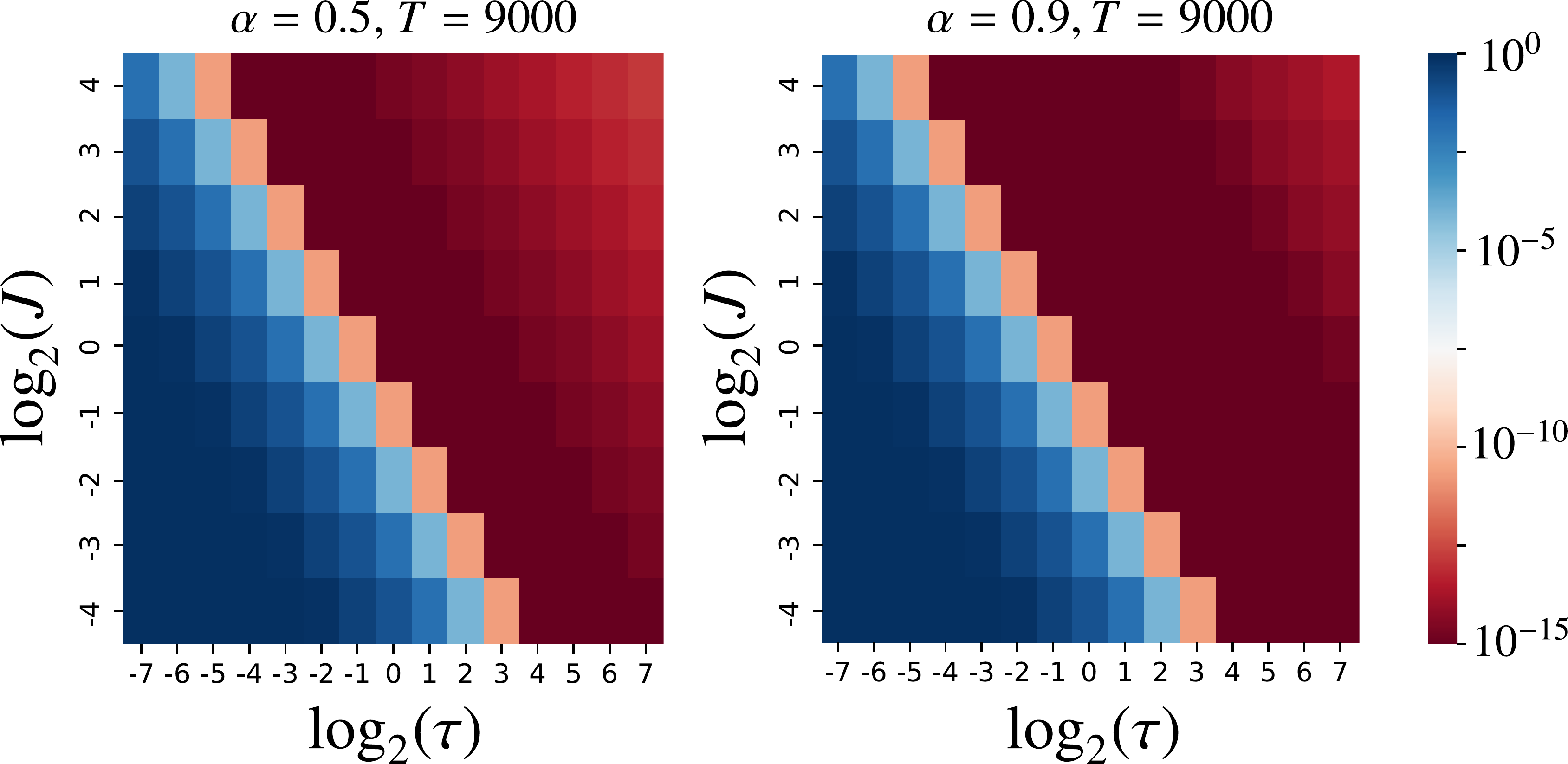}
	\caption{The average QESP indexes with $\alpha=0.5$ (left), and $\alpha=0.9$ (right) along the variation of $J$ and $\tau$ over 10 random trials for the HQR-5 of five qubits, $V=1$. The transient time steps and the evaluation time steps are $9000$ and $1000$, respectively.
	}\label{fig:esp}
\end{figure}

\begin{figure}
	\centering
	\includegraphics[width=1.0\linewidth]{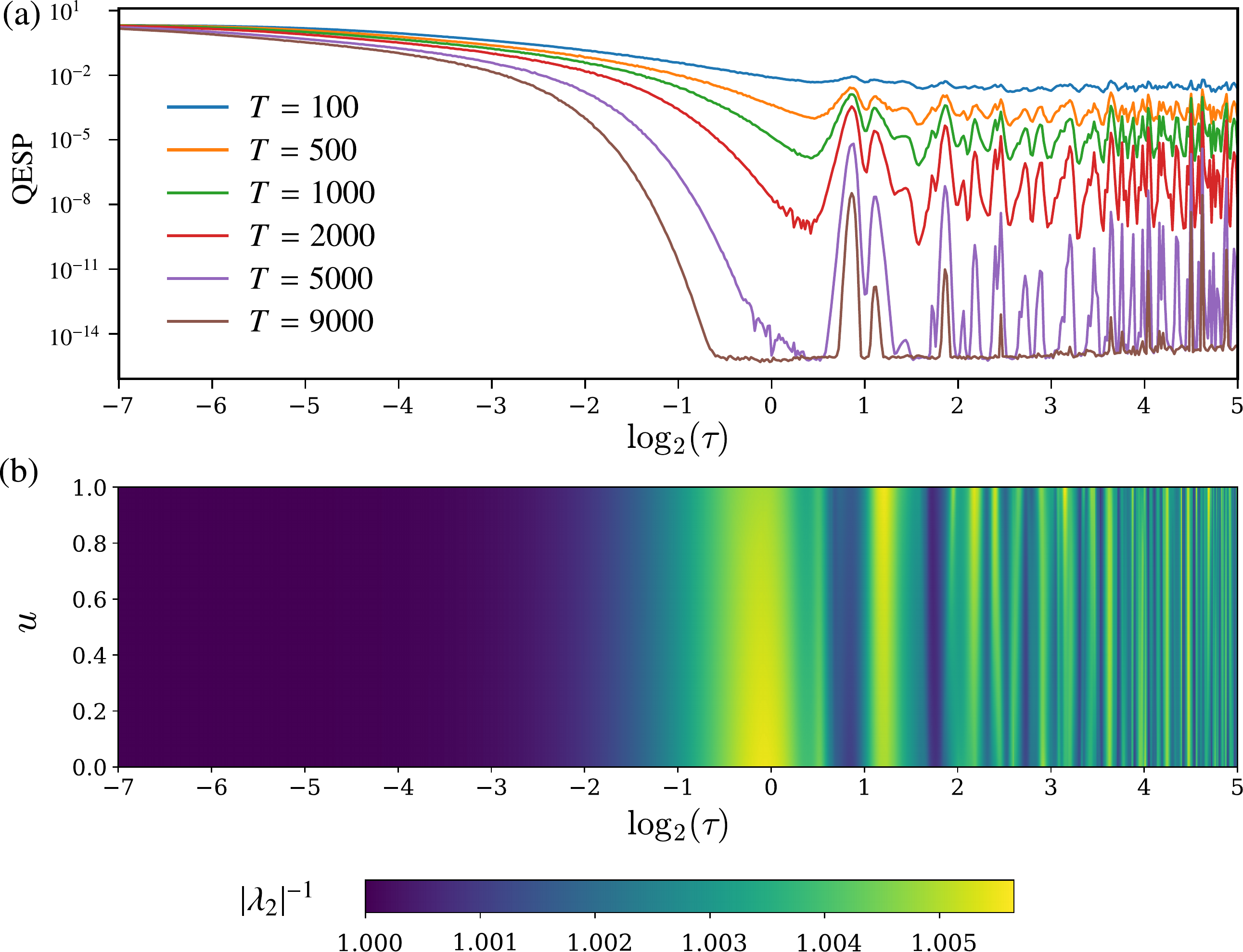}
	\caption{(a) The average QESP indexes with $\alpha=0.5, V=1,$ and $J=1.0$ along the variation of $\tau$ over 10 random trials for the HQR-5 of five qubits. These values are evaluated in the dynamics range of $[T, T+1000]$ with different values of $T$.
	(b) The color map of $|\lambda_2|^{-1}$  for a single QR of five qubits, where $\lambda_2$ is the second largest eigenvalue in magnitude of the CPTP map with the constant input $u\in[0,1]$ and input interval $\tau\in[2^{-7}, 2^5]$.
	}\label{fig:spec}
\end{figure}

\begin{figure}
	\centering
	\includegraphics[width=1.0\linewidth]{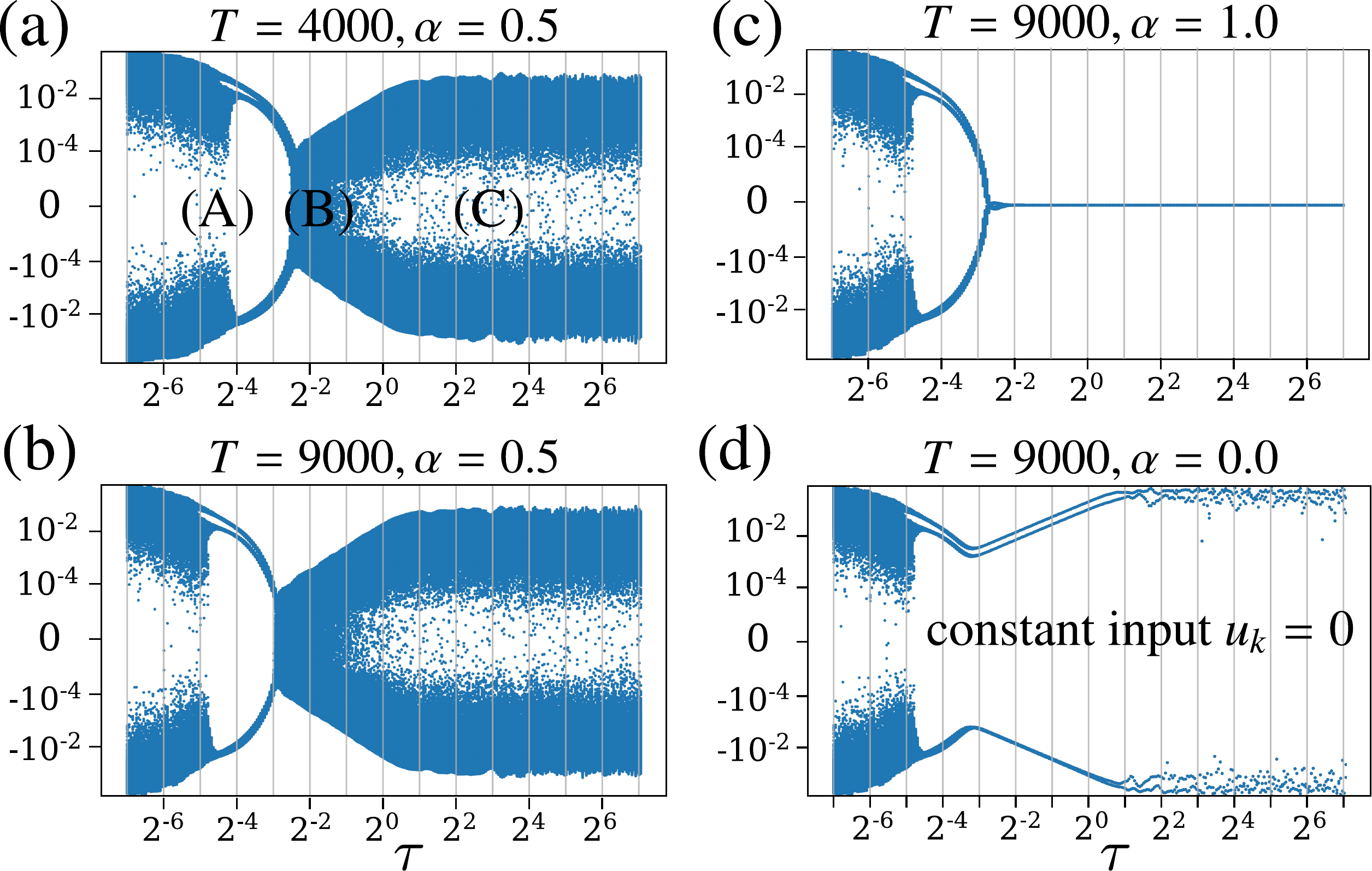}
	\caption{The overlaid values of four representative signals in one QR (excluding the signals from the input qubit) for the range of dynamics in $[T, T+1000]$. The diagrams are shown with different setting of
		$T, \alpha$ and different types of the input. (a) Random input, $T=4000, \alpha=0.5$. (b) Random input, $T=9000,\alpha=0.5$. (c) Random input, $T=9000,\alpha=1.0$ (withdraw the external input). (d) Constant input without the feedback ($T=9000, \alpha=0.0$).
		The behavior of signals are divided into three regimes as illustrated in Fig.~\ref{fig:bifu}(a): (A) the dynamics approach fixed points for long enough transient time $T$, (C) the dynamics exhibits two separated regions in responding to the external inputs, and (B) the transition between (A) and (C).
	}\label{fig:bifu}
\end{figure}

Figure~\ref{fig:esp} presents the average QESP indexes along the variation of $J$ and $\tau$ over 10 random trials for the HQR-5 of five qubits, $V=1$, $\alpha=0.5, 0.9$, for the range of dynamics in $[T, T+E]$ ($T=9000, E=1000$ time steps) with $P=10$.
The empirical QESP validity is characterized by a border of a sharp stable-unstable transition.
The theoretical reason for this behavior can be understood as the influence of the initial state in the relaxation dynamics, which is explained as follows.
We consider an arbitrary operator $\hat{A}$ applied to a single QR between two consecutive inputs $u_k$ and $u_{k+1}$.
Since the QR continues evolving for a time interval $\tau$,
the interaction representation of the operator in this interval is $\hat{A}(t) = \ehm \hat{A}(0)\ehp$ ($0\leq t\leq \tau$),
where $H$ is the Hamiltonian of the QR.
The auto correlation function $G(\tau)=\langle \hat{A}(0)\hat{A}(\tau) \rangle$ at the absolute zero temperature can be written 
in the following form~\cite{dutta:2015:ising}:
\begin{align}\label{eq:corr}
G(\tau) = \sum_m e^{-(E_m-E_0)\tau}| \langle 0 | \hat{A}(0)|m \rangle|^2,
\end{align}
where $\{\ket{m}\}$ are the eigenstates of $H$, which correspond with the energies $\{E_m\}$.
Here, $E_0$ is the lowest energy, that is, the smallest eigenvalue of $H$.

Equation~\eqref{eq:corr} implies that the time auto correlation function will decay exponentially with the dominant term $\exp(-\tau\Delta_J)=\exp(-J\tau\Delta)$, where $\Delta_J =  J\Delta$ is the smallest nonzero eigengap of $H$~\cite{dutta:2015:ising}.
Here, $\Delta$ is the smallest nonzero eigengap of $H/J$ (in our Hamiltonian dynamic, changing $J$ does not affect the value of $\Delta$).
If $J\tau$ is too small, the dynamics approach an identity map between two consecutive inputs, the influence of the initial state of the QR system therefore remains during a long time step. 
If we increase $J\tau$, the dynamics relax exponentially quickly between inputs, thereby decreasing the effects of the initial state on the system.
Furthermore, as shown in Fig.~\ref{fig:esp}, increasing the connection strength $\alpha$ (from $0.5$ to $0.9$) will lead to a slight decrease in the values of the QESP index, which can reduce the number of time steps in the washout phase.
It is interesting to note that if we fix $J$, the value of the QESP index does not simply decrease as we increase $\tau$.
We can verify this observation when plotting the values of the QESP index along the variation of $\tau$ with $J=1.0$ and different numbers of time steps $T$.
These values are evaluated for the range of dynamics in $[T, T+1000]$.
We can explain this behavior via the spectral analysis of the CPTP map, which evolves the density matrix of each QR system. 

\begin{figure*}
	\includegraphics[width=1.0\linewidth]{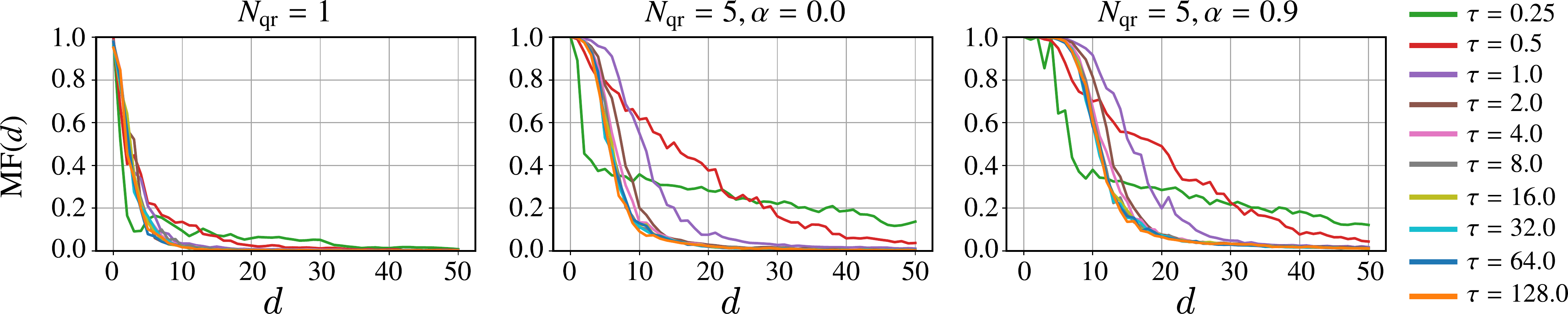}
	\caption{The memory functions $\mf(d)$ of the HQR system according to delay $d$ with different settings $\nqrc, \alpha,$ and $\tau$. The memory functions are evaluated as the average functions on 10 samples of HQR with respect to random coupling.}
	\label{fig:memfunc}
\end{figure*}

\begin{figure*}
 \includegraphics[width=1.0\linewidth]{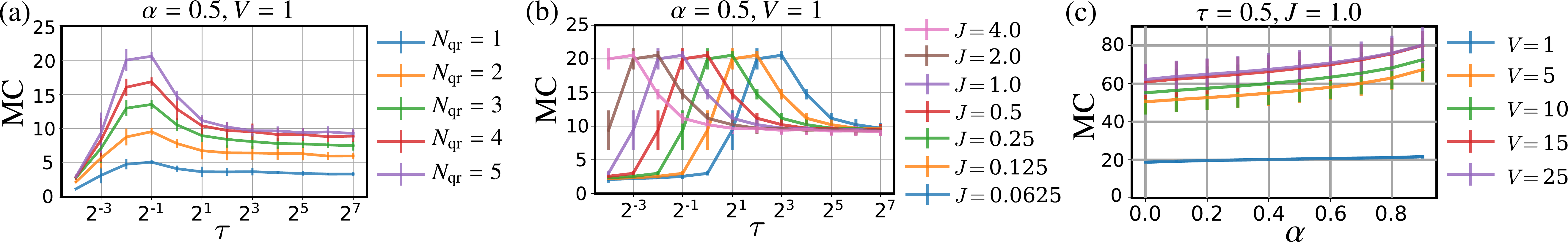}
  \caption{Memory capacity shown as a function of the time interval $\tau$ for the HQR system, which is formed by $\nqrc$ QR systems of five qubits. The number of time steps used in the washout, training, and evaluation phases are 1000, 3000, and 1000, respectively. The memory capacity is calculated for delay $d=0,1,\ldots,200$, and is averaged over 10 random trials of different runs. For all the plots, the error bars indicate the standard deviations. (a) $\alpha=0.5, V=1, J=1.0$, (b) $\alpha=0.5, V=1, \nqrc=5$, (c) $\alpha=0.5, J=1.0, \nqrc=5$.}
  \label{fig:memcapa:V1}
\end{figure*}

We consider a single QR case to simplify our theoretical explanation. From Eq.~\eqref{eqn:input:tr}, we can rewrite the evolution of the density matrix of the entire system from $t=(k-1)\tau$ to $t=k\tau$ via a CPTP map $\cL_k$, such that $\rho^{(k)} = \cL_k(\rho^{(k-1)})$, where $\rho^{(k)}$ is the density matrix of the QR at $t=k\tau$.
Since the state $\rho^{(k)}$ depends only on the state $\rho^{(k-1)}$ and the $k$th input, the evolution is Markovian.
We therefore have $\rho^{(k)}=\cL_k\circ \cL_{k-1}\circ \ldots \circ \cL_1(\rho^{(0)})$.
Since a CPTP map is a contraction map, with respect to the Schatten 1-norm, the density matrices satisfy decreasing system distinguishability $\Vert \rho^{(k)} - \hat{\rho}^{(k)}\Vert_1 \leq \Vert \rho^{(k-1)} - \hat{\rho}^{(k-1)}\Vert_1$~\cite{nielsen:2011:QCQI}.
Any density matrix $\rho$ can be converted into vector form by stacking columns of $\rho$, which we denote as $\vc(\rho)$. This conversion allows us to define a linear space of matrices (Fock-Liouville space) associated with an inner product.
The map $\cL_k$ can now be expressed as a matrix $\tilde{\cL}_k$, such that $\vc(\rho^{(k)})=\tilde{\cL}_k\vc(\rho^{(k-1)})$.

In the case of the one-dimensional constant input $u_k=u\in[0,1]$, we have $\mathcal{L}_k\equiv \mathcal{L}$ for all $k$ and $\rho^{(k)}=\mathcal{L}^{k}(\rho^{(0)})$.
The convergence property of $\rho^{(k)}$ can then be explored via the spectrum of CPTP map $\mathcal{L}$~\cite{bruneau:2014:repeated}.
The spectrum of $\cL$ can be written as $1=|\lambda_1|\geq |\lambda_2| \geq \ldots \geq |\lambda_s|$, where $\lambda_j$ is the $j$th eigenvalue of $\tilde{\cL}$. Because $\cL$ is not Hermitian, it has both left and right eigenmatrices. We denote $L_j$ and $R_j$ as the vector forms of the left and right eigenmatrices corresponding to eigenvalue $\lambda_j$, respectively.
Since $\vc(\rho^{(k)}) = \tilde{\cL}^k(\vc(\rho^{(0)}))$, we have
\begin{align}
    \vc(\rho^{(k)}) = \vc(R_1) + \sum_{j=2}^sc_j\lambda_j^k\vc(R_j),
\end{align}
where $c_j = \vc(L_j)^{\dagger}\vc(\rho^{(0)})$.
Therefore, if $|\lambda_2|^{-1} > 1$, the steady state of the system is $R_1$, and the convergence rate depends on the magnitude of $|\lambda_2|^{-1}$.
The QESP index of the system will be reduced more quickly with a larger value of $|\lambda_2|^{-1}$.
Figure~\ref{fig:spec}(b) presents the color map of $|\lambda_2|^{-1}$ for variations of the constant input $u\in[0, 1]$ and the input interval $\tau\in[2^{-7}, 2^5]$.
The yellow parts of Fig.~\ref{fig:spec}(b) correspond with the values of $\tau$ for a higher $|\lambda_2|^{-1}$. 
Interestingly, these values of $\tau$ also lead to a lower value of the QESP index.
Figure~\ref{fig:spec}(b) therefore presents an indicator to set the value of $\tau$ and the number of transient time steps, such that our HQR system can achieve the QESP after the washout phase. 

We further look into the dynamics of the HQR-5 via the bifurcation diagrams where we consider $\tau$ is the driving parameter, $J=1.0$, and $V=1$.
In Fig.~\ref{fig:bifu}, the values of four representative signals in one QR (excluding the signals from the input qubit) are overlaid for the range of dynamics in $[T, T+1000]$.
In Fig.~\ref{fig:bifu}(a), we consider the diagram at $\alpha=0.5$ as an example to explore three intriguing regimes: (A) the dynamics approach fixed points for long enough transient time $T$, (C) the dynamics exhibits two separated regions in responding to the external inputs, and (B) the transition region between (A) and (C).
We note that the left part of (A) appears as a broad band even if the external input is withdrawn ($\alpha=1.0$)~[Fig.~\ref{fig:bifu}(c)],
or setting the constant inputs without the feedback ($\alpha=0.0$)~[Fig.~\ref{fig:bifu}(d)].
Therefore, this part remains as the effect of the initial state during the short transient time $T$,
which also corresponds with the dark side in Fig.~\ref{fig:spec}(b).
As we increase $T$, the region of $\tau$ in (B) will extend to the left~[Fig.~\ref{fig:bifu}(b)]. 
Interestingly, (B) corresponds to the unstable-stable border in Fig.~\ref{fig:esp}. 
Therefore, setting $\tau$ at the end of (B) or the beginning of (C) can avoid the effect of the initial state and lead to the optimal trade-off for the effect of the past input and the relaxation dynamics, which may give a high ability to memorize the past patterns.

\paragraph{Memory capacity (MC). }

\begin{figure*}
  \centering
\includegraphics[width=1.0\linewidth]{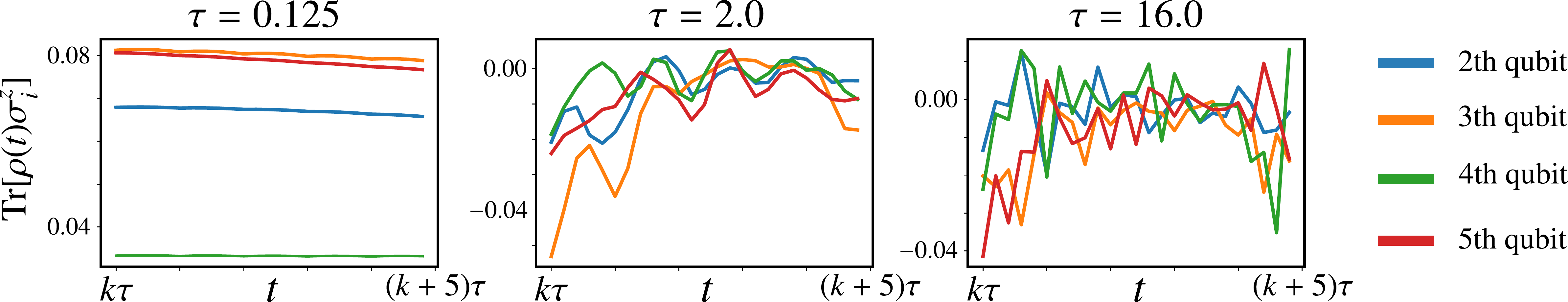}
  \caption{The signals $\tr[\rho(t)\sigma_i^z]$ obtained from $i$th qubit (excluding the signals from the input qubits) in one QR system for window length $5\tau$ of HQR-5 with the connection strength $\alpha=0.5$ and the coupling magnitude $J=1.0$.}
  \label{fig:signal}
\end{figure*}

The property of the HQR in storing information of recent inputs to the current states is commonly measured by memory capacity $\mc = \sum_{d=0}^{\infty}\mf(d)$, where $\mf(d)$ is the memory function that evaluates the capacity to reconstruct the previous $d$ steps of the input~\cite{jaeger:2001:short}.
This implies that if we set the input as a random sequence $\{u_k\}$ in $[0, 1]$, then the forecast and target are $y_k$ and $\hat{y}_k=u_{k-d}$, respectively.
$\mf(d)$ is then defined as $\mf(d)=\dfrac{\textup{cov}^2(y_k, \hat{y}_k)}{\sigma^2(y_k)\sigma^2(\hat{y}_k)} \in [0, 1]$,
where $\textup{cov}(x,y)$ and $\sigma(x)$ express the covariance and the standard deviation, respectively.
A dynamical system with high MC is superior for temporal forecasting tasks that need to utilize historical patterns.

In Fig.~\ref{fig:memfunc}, $\mf(d)$ is plotted as a function of $d$ for the HQR system, which contains $\nqrc$ QR systems. 
We consider the coupling magnitude $J=1.0$ in the Hamiltonian, the number of virtual nodes $V=1$ in each QR system, the number of QR systems $\nqrc = 1, 5$, and the connection strength $\alpha=0.0, 0.9$ with different settings of the time interval $\tau$ to inject the input.
Here, each QR has five qubits and the number of time steps used in the washout, training, and evaluation phases are 1000, 3000, and 1000, respectively.
The memory function is evaluated on 10 samples of HQR with respect to the random coupling.
It is evident that increasing $\nqrc$ will increase the value of $\mf(d)$ for any $\tau$, thereby indicating the relatively large value of $\mf(d)$ in the region of the large delay $d$.
For the HQR-5 system, the connection strength $\alpha=0$ implies that we have a spatial multiplexing setup where all QR systems are disjoint.
By introducing the feedback, for example, $\alpha=0.9$, the values of $\mf(d)$ are higher in the region of the small delay, $d$.

Figure~\ref{fig:memcapa:V1} shows MC as functions of
$\tau$ for the HQR along with different $\alpha$, $\nqrc$ [Fig.~\ref{fig:memcapa:V1}(a)], $J$ [Fig.~\ref{fig:memcapa:V1}(b)], and $V$ [Fig.~\ref{fig:memcapa:V1}(c)].
MC is calculated for delay $d=0,1,\ldots,200$,
and is averaged over 10 different runs with random trials of Hamiltonian coefficients and connection coefficients in the HQR.
We observe from Fig.~\ref{fig:memcapa:V1}(a) and Fig.~\ref{fig:memcapa:V1}(c) that increasing $\nqrc$ and $V$ will lead to an increase in the MC of the system, while the MC becomes saturated around $V=15$ [Fig.~\ref{fig:memcapa:V1}(c)].
The connection strength $\alpha$ measures the dominant role of the reservoir states in the feedback,
which increases the ability of memorizing past patterns.
As shown in Fig.~\ref{fig:memcapa:V1}, increasing $\alpha$ while keeping the external input ($\alpha < 1$) enables the extensive MC.
Furthermore, for each $J$, there exists an optimal $\tau$ to maximize the MC,
which is close to the border of  unstable-stable transition in Fig.~\ref{fig:esp}.
The reason for this behavior can be understood as a trade-off between the influence of the input and the effect of the relaxation dynamics in the QR.
As shown in Fig.~\ref{fig:signal}, if $\tau$ is too small, the dynamics approach an identity map, therefore the signals are almost linear without prediction power. If $\tau$ is too large, the signals seem to fluctuate in the interval $\tau$ because the dynamics relax exponentially fast between inputs, thereby decreasing the effects of past inputs into the system.

\subsection{Quantum-innate training with noise robustness\label{sec:innate:noise}}

In the current HQRC framework, the pattern produced by each QR can be affected by statistical error noise in the observed signals.
If the QRs are connected via mutual feedback in the ensemble system,  
the changes resulting from noise can repeatedly propagate along the feedback pathway and affect the system's activities.
Here, the linear feedback scheme in the HQRC framework allows us to reduce the effect of this noise by introducing the \textit{quantum-innate training} scheme, which resembles the innate training scheme used in the classical RC framework~\cite{laje:2013:innate}.
If we consider our HQR's trajectories in the absence of noise to be the innate trajectories,
we can tune the linear feedback connection $\wfeed$ to adapt the patterns with noise, returning them to the stable innate trajectories. 
We briefly describe the training process according to the first-order reduced and controlled error (FORCE) learning algorithm~\cite{sussillo:2009:FORCE}.
Before the training, the innate target trajectories are recorded by letting the HQR evolve under the  input and configuration used for each QR during training.
In the training procedure, the HQR evolves according to algorithm~\ref{alg:hqrc} with the presence of noise and random initial conditions for a number of training loops.
Let $w_{ni}$ denote the connection weight from node $i$ ($i=1,\ldots,\ntot)$ of the HQR to the $n$th QR ($n=1,\ldots,\nqrc)$. The update for $w_{ni}$ is
\begin{align}
    w_{ni}(t) = w_{ni}(t-\Delta t) - \zeta e_n(t) \sum_{j\in A(n)}P^{(n)}_{ij}r_j(t),
\end{align}
where $A(n)$ is the list of all nodes connected to the $n$th QR, $r_j(t)$ is the trajectory of the $j$th node in HQR, and $\zeta$ is the learning rate that adjusts the amount of the update.
Here, $e_n(t)$ represents the difference in the activity of the $n$th QR before and after the update. For example, we can choose the representative spin in $n$th QR to consider its trajectory and compute $e_n(t)$ based on the difference between this trajectory and the innate target trajectory.
Furthermore, for each $n=1,\ldots,\nqrc$, we define the matrix $\bP^{(n)}$ with its elements $P^{(n)}_{ij}$, where $\bP^{(n)}$ is a square matrix that estimates the inverse of the correlation matrix of the nodes in HQR that connect to the $n$th QR. The value of $P^{(n)}_{ij}$ is updated as follows:
\begin{align}
    P_{ij}^{(n)}&(t)=P_{ij}^{(n)}(t-\Delta t)-\\
    &\frac{\sum_{k \in A(n)} \sum_{l \in A(n)} P_{ik}^{(n)}(t-\Delta t) r_{k}(t) r_{l}(t) P_{lj}^{(n)}(t-\Delta t)}{1+\sum_{k \in A(n)} \sum_{l \in A(n)} r_{k}(t) P_{kl}^{(n)}(t-\Delta t) r_{l}(t)}.\nonumber
\end{align}

\begin{figure*}
    \includegraphics[width=1.0\linewidth]{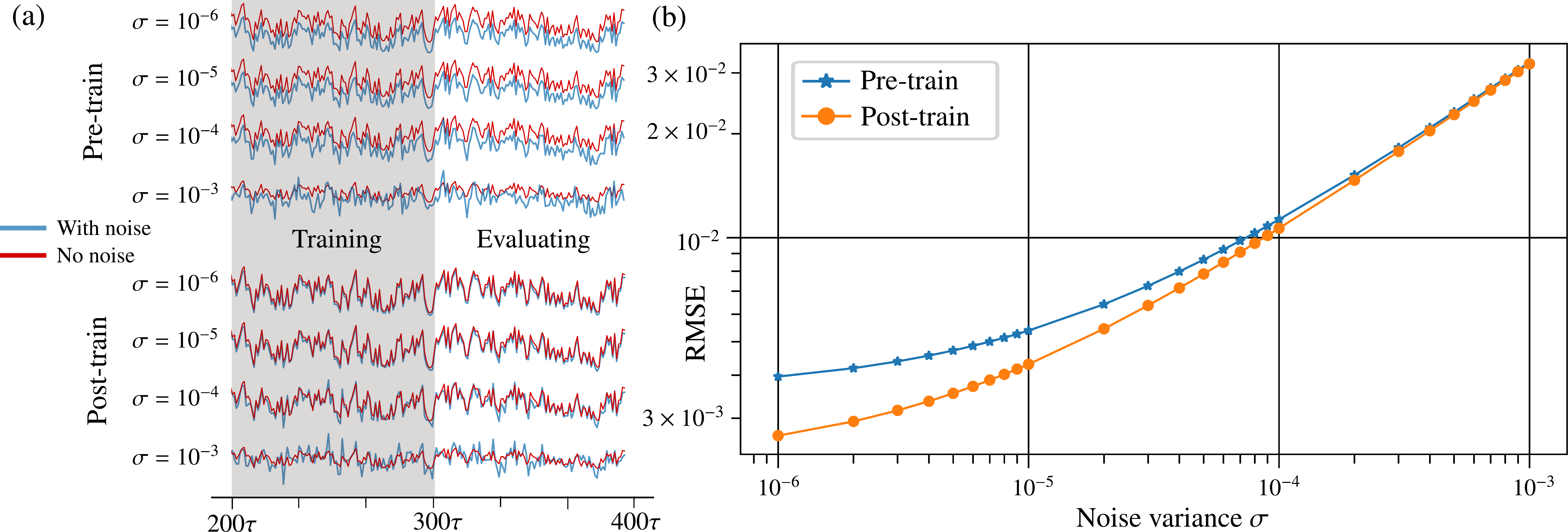}
	\caption{(a) The representative activities in the HQR at four different levels of noise in pre-training and post-training after 20 innate training loops. Red indicates the target trajectory (no noise), and blue indicates the test trajectory (with noise). 
	(b) Root mean square error (RMSE) metric of the evaluating trajectories and the target trajectories in pre-train and post-train regimes according to different levels of noise. The metric is averaged over 10 random trials of HQR. }
	\label{fig:denoise}
\end{figure*}

\begin{figure*}
  \centering
  \includegraphics[width=1.0\linewidth]{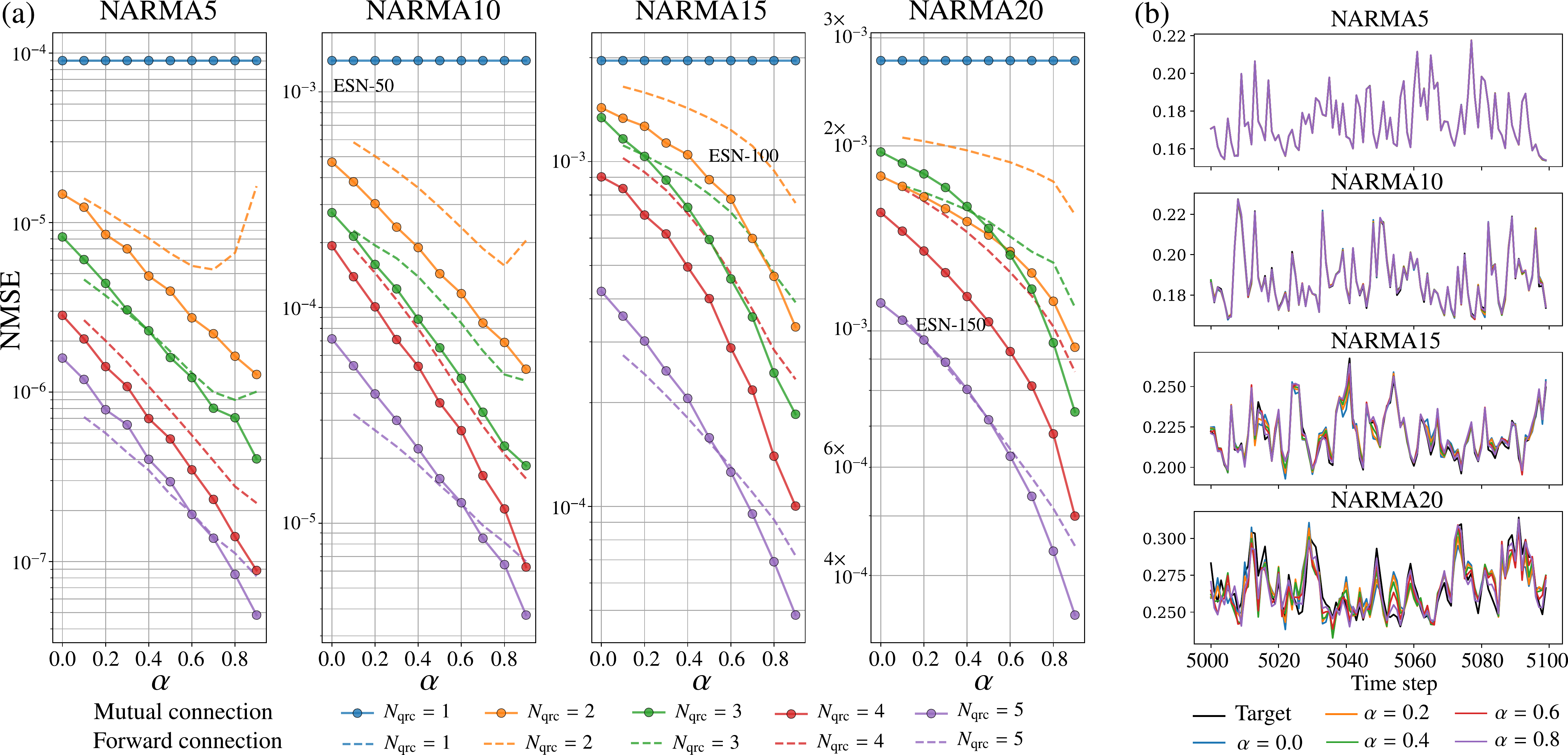}
  \caption{(a) The averaged NMSE for the NARMA tasks according to the number $\nqrc$ of QR systems in HQR and the connection strength $\alpha$. Each plot of NARMA10, NARMA15, and NARMA20, contains the performance at $\textup{NMSE} \approx 10^{-3}$ for the conventional echo state network (ESN)~\cite{jaeger:2004:rc} such as, for example, ESN-100 for the ESN with 100 nodes.
  (b) Typical output time series for the NARMA tasks of the HQR-5. Other parameters for (a)(b) are $V=20, J=1.0,$ and $\tau=1.0$.}\label{fig:narma}
\end{figure*}

We demonstrate our quantum-innate training scheme in the following experiment. We use an HQR comprising five QRs with the number of virtual nodes $V=1$, the connection strength $\alpha=0.5$, $J=1.0$, and  $\tau=4.0$. We introduce the continuous Gaussian noise with a mean of zero and a variance of $\sigma$ into the reservoir signals at each time step. Due to the feedback scheme, without the innate training, the trajectory deviates from the target trajectory with the accumulation of noise.
In Fig.~\ref{fig:denoise}, we present the representative deviated trajectories of our HQR, corresponding with four different levels of noise as $\sigma=10^{-6}, 10^{-5}, 10^{-4}, 10^{-3}$. 
The training is performed on the training window $[t_{\textup{transient}}:t_{\textup{train}})$ and is evaluated on the evaluating window $[t_{\textup{train}}:t_{\textup{eval}})$.
Figure~\ref{fig:denoise}(a) demonstrates that after 20 training loops (post-train) with learning rate $\zeta=10$, the deviated trajectories with noise (pre-train) are returned well to the target trajectories in both of the training and evaluating window. 
Therefore, the innate training procedure can solve the problem of noise sensitivity originated from the feedback connection in HQR system.
Figure~\ref{fig:denoise}(a) shows the average root mean square error (RMSE) metric between the evaluating trajectories and the target trajectories in the pre-train and post-train regimes over 10 random trials of HQR.
Here, we set $t_{\textup{transient}}=2000\tau, t_{\textup{train}}=4000\tau$, and $ t_{\textup{eval}}=6000\tau$.

\section{Benchmarks}
\subsection{NARMA tasks}

We first demonstrate the performance of the HQRC in the NARMA benchmark~\cite{atiya:2000:narma},
which is commonly used for evaluating the computational capability for temporal processing with long time dependence.
The NARMA system is formulated as the $n$th-order nonlinear dynamical system, which has the following form: $y_k = \kappa y_{k-1} + \eta y_{k-1}\left( \sum_{j=0}^{n-1}y_{k-j-1}\right) + \gamma u_{k-n+1}u_k + \delta$,
where $\kappa=0.3, \eta=0.05, \gamma = 1.5,$ and $\delta=0.1$ in our experiments.
We consider $n=5, 10, 15, 20,$ where the corresponding systems are NARMA5, NARMA10, NARMA15, and NARMA20, respectively.
To set $y_k$ into the stable range, we linearly scale the input $u_k$ from [0, 1] to [0, 0.2].
The number of time steps are set as 2000, 2000, and 2000 for the washout, the training, and the evaluation phase, respectively.
The performance is evaluated using the normalized mean-squared error (NMSE) metric,
\begin{align}
\textup{NMSE} = \frac{\sum_{k=4001}^{6000} (y_k-\hat{y}_k)^2}{\sum_{k=4001}^{6000} \hat{y}^2_k},
\end{align}
where $y_k$ and $\hat{y}_k$ are the prediction and the target of system at time step $k$.
We present two types of connections in HQR: the mutual connection where any two QRs are mutually connected, and the forward connection where each QR only connects with the next QR in a forward direction.
Figure~\ref{fig:narma}(a) illustrates the averaged NMSE over 10 random trials for the HQRs comprising $\nqrc=1\text{--}5$ QRs with five qubits, $V=20$, and $\alpha=0.0\text{--}0.9$,
where $\alpha=0.0$ corresponds with spatial multiplexing scheme presented in Ref.~\cite{nakajima:2019:qrc}. 
We set $J$ and $\tau$ in the stable region of the QESP index such as $J=1.0$ and $\tau=1.0$.
These values are close to the stable-unstable border in Fig.~\ref{fig:esp} to increase the MC, but at the beginning of (C) as shown in Fig.~\ref{fig:bifu}(a) to avoid the effect of the initial state. 
Because higher-order NARMA tasks require high memory to predict, increasing $\alpha$ will boost the performance for both types of connections.
As a reference, each plot of NARMA10, NARMA15, and NARMA20, contains the performance at $\textup{NMSE} \approx 10^{-3}$ for the conventional echo state network (ESN)~\cite{jaeger:2004:rc} such as, for example, ESN-100 for the ESN with 100 nodes.
Figure~\ref{fig:narma}(b) illustrates the typical outputs for the HQR-5 in the evaluation phase. 
The outputs with higher $\alpha$ fit well to the targets for all tasks, even for a difficult task like NARMA20.


\subsection{Emulating chaotic systems\label{sec:chaos}}

\begin{figure*}
	\centering
	\includegraphics[width=1.0\linewidth]{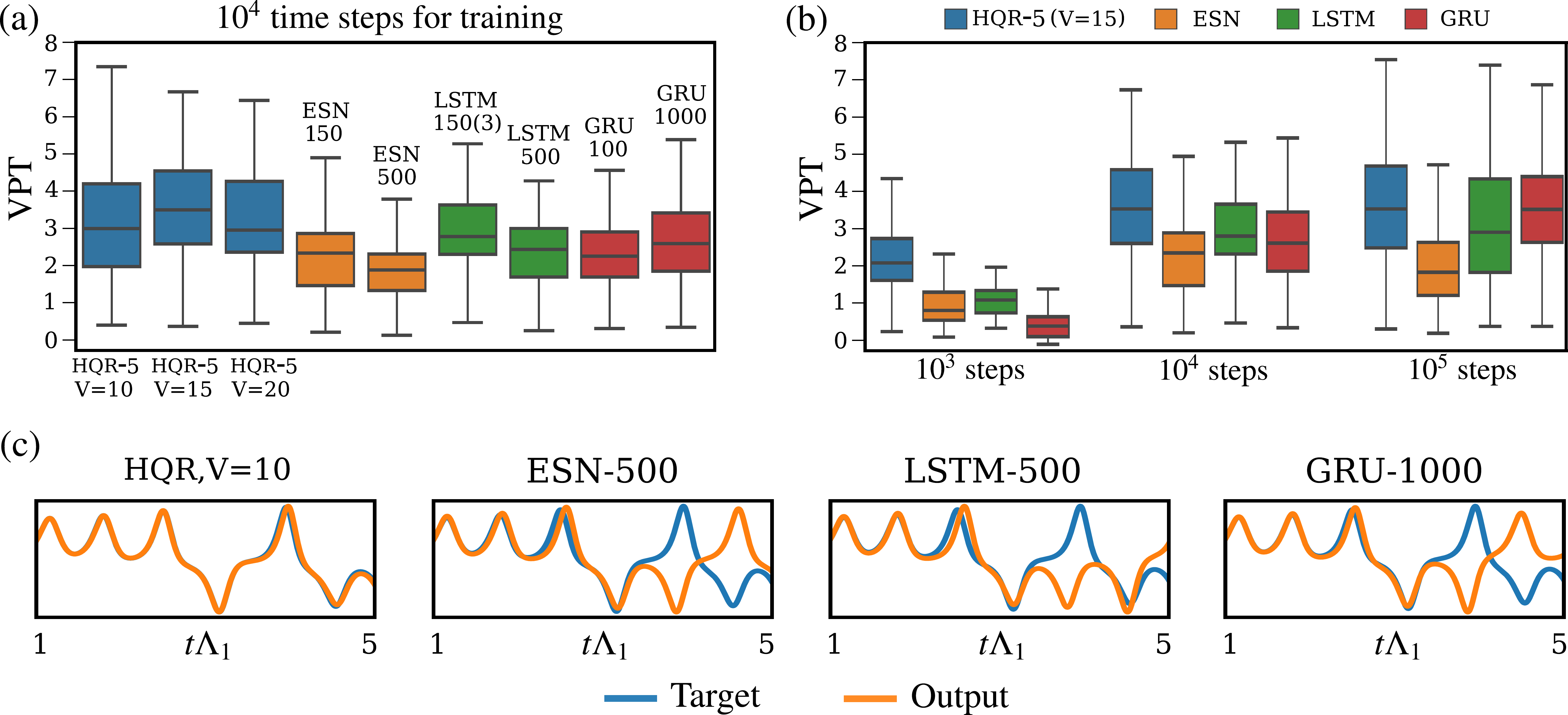}
	\caption{(a) Box plots displaying the distribution of the VPT over 100 random predictions with $10^4$ time steps for training. The interquartile range is contained
		within the box, and the 5th and 95th percentiles are marked by the whiskers. 
		The median is the line across the box. (b) The performance of prediction models due to the number of time steps for training.
		(c)  Certain typical predicted time series on the Lorenz system when $10^4$ time steps are used for training. The time-scale of time series is normalized with Lyapunov time $\Lambda_1^{-1}$, 
		where the maximum Lyapunov exponent $\Lambda_1$ of the Lorenz attractor system in our study is $\Lambda_1=0.9056$~\cite{viswanath:1998:lyap}.
		}\label{fig:lorenz:chaos}
\end{figure*}

Next, we demonstrate the performance of the HQRC on emulating chaotic systems, particularly for high-dimensional input that the normal QRC is unable to implement.
In this task, the system learns the input of the next step.
After learning the readout weights to fit the targets in the stage called teacher forcing stage, 
the external input streams are withdrawn and the output is fed into the input so that the system is closed.
Thereafter, the system can automatically evolve and replicate the dynamical evolution of the target system but within a relatively short prediction horizon because of the exponential divergence between the trajectories of the reservoir and the target system.
We define the normalized root mean square error 
$
    \nrmse(\by_t) = \sqrt{ \frac{1}{M}\sum_i\frac{(y_{ti} - \hat{y}_{ti})^2}{\hat{\sigma}_i^2}},
$
where $\by_t=(y_{ti})_i$ and $\hat{\by}_t=(\hat{y}_{ti})_i$ are the $M$-dimensional forecast and target, and $\hat{\sigma}_i$ is the standard deviation of the target in time of each component $i$.
As referenced from Ref.~\cite{vlachas:2020:backpropagation}, 
to evaluate the prediction performance, we compute the valid prediction time $
    \vpt = \Lambda^{-1}_1\textup{argmax}_{t_f}\{\nrmse(\by_t) \leq \varepsilon, \forall t\leq t_f\}
$, which is the largest time $t_f$ (normalized with respect to the maximum Lyapunov exponent $\Lambda_1$ of the chaotic system) where the NRMSE error is smaller than $\varepsilon$ ($\varepsilon=0.5$ in our experiments).
Large VPT means long prediction horizon in the performance of the model.

We employ two chaotic systems: the Lorenz attractor and the Kuramoto-Sivashinsky equation (KSE)~\cite{sivas:1977:ks,kuramoto:1978:diff} with spatiotemporally chaotic solutions.
The Lorenz attractor is given by three ordinary differential equations: $dx/dt=a(y-x), dy/dt=x(b-z)-y, dz/dt=xy-cz,$ where $(a, b, c)=(10, 28, 8/3)$.
The time series is obtained by the fourth-order Runge-Kutta method with step size $\Delta t=0.01$, and the model attempts to predict $x(t+\Delta t)$ from $x(t)$.
The first $K=10^4$ steps are used for training, and $10^3$ steps are predicted iteratively.
The input signals are min-max scaled to be in the range $[0.0, 1.0]$.
Our HQR-5 of six qubits, with $\tau=4.0, J=2.0$, and $V\in\{10,15,20\}$, corresponds with $6\times 5\times V$ computational nodes.
For chaos emulating tasks, the reservoir needs to learn an arbitrarily good approximation of the chaotic recurrence. The non-linear transformation is required, but adding it will degrade the MC~\cite{ganguli:2008:memory},
thus, we need careful considerations on this memory nonlinear trade-off of the reservoir.
In this context, $\alpha$ adjusts this trade-off, which is tuned for the best performance in our experiments.

We adopt the framework in \cite{vlachas:2020:backpropagation} to generate the simulation data and to implement the common ML models such as long short-term memory (LSTM)~\cite{hochreiter:1977:LSTM}, gated recurrent unit (GRU)~\cite{choetal:2014:learning}, and echo state network (ESN)~\cite{jaeger:2004:rc}--for example, LSTM-$n$($m$) for the LSTM with $n$ hidden units and $m$ layers (if $m>1$), and ESN-$n$ for the ESN with $n$ nodes, where
$n \in \{80,100,120,150,500,1000,1500,3000\}$.
We set the number of layers in the LSTM and GRU in $\{1,2, 3\}$.
The box plots in Fig.~\ref{fig:lorenz:chaos}(a) demonstrate the distribution of the VPT over 100 random tests for the HQR model, and the two best results in each of the other models. HQR-5 shows superior performance, as its VPTs are highest, while other models cannot capture long time steps in the prediction as shown in the typical examples placed Fig.~\ref{fig:lorenz:chaos}(c).
As demonstrated in Fig.~\ref{fig:lorenz:chaos}(b), for different training time steps $K$,
the HQR are still better even with a small number of the computational nodes ($V=15$ corresponds with 450 nodes).
The best results of the ESN, LSTM, and GRU models are selected to plot in Fig.~\ref{fig:lorenz:chaos}(b).

\begin{figure*}
	\centering
	\includegraphics[width=1.0\linewidth]{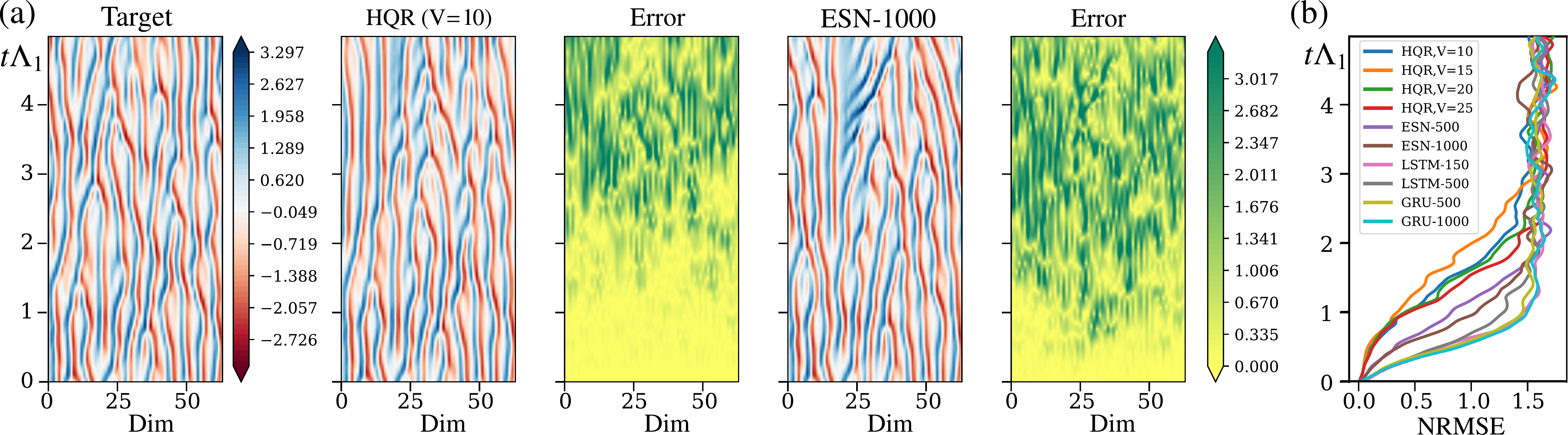}
	\caption{(a) Contour plots of a spatio-temporal forecast on the KSE with the absolute difference (Error) between the target and the prediction for the parallel HQR and parallel ESN. (b) The evolution of the NRMSE averaged over 10 random predictions on the KSE. The time-scale in each plot is normalized with Lyapunov time $\Lambda_1^{-1}$, where the maximum Lyapunov exponent $\Lambda_1$ of the KSE in our study is $\Lambda_1=0.05$~\cite{fan:2020:longterm}.}\label{fig:ks:heat}
\end{figure*}

The KSE is the partial differential equation $\frac{\partial u}{\partial t} + u\frac{\partial u}{\partial x} + \frac{\partial^2 u}{\partial x^2} + \frac{\partial^4 u}{\partial x^4}=0$ for the scalar function $u(x,t)$ 
in the interval $x\in [0, L)$ 
with periodic boundary conditions $u(x,t)=u(x+L,t)$.
The KSE is integrated on a spatial domain of $M=64$ uniform subintervals with $\Delta t = 0.25, L=22$, thereby yielding a simulated data set with $M$-dimensional time series, which can be used as multi-dimensional input-target pairs.
The simulation is performed using the fourth-order method for stiff PDEs~\cite{kassam:2005}. We utilize the framework in Ref.~\cite{vlachas:2020:backpropagation} to simulate the solution of KSE up to $T=6\times 10^4$, which corresponds to $24\times 10^4$ samples.
The first $4\times 10^4$ are excluded for initial transients, and the remaining data are divided into training and testing dataset of $10^5$ samples in each dataset.
For the HQR and the ESN model, we employ the augmentation technique proposed in Ref.~\cite{pathak:2018:prl}.
Here, the hidden states are augmented such that the hidden states are squared in half of the computational nodes.

We employ the parallel architecture in Ref.~\cite{pathak:2018:prl} to build $32$ HQRs,
where each HQR predicts a spatially 2-dimensional local region.
We divide the input $\bu_k$ into $32$-local groups $\bu^{(i)}_k$, where $\bu^{(i)}_k$ comprises the $i$th and $(i+1)$th element in $\bu_k$.
The $i$th HQR is the HQR-10 with six qubits to receive $10$-dimensional inputs to predict $\bu^{(i)}_{k+1}$.
The input for each HQR is the concatenated vector of $\bu^{(j)}_k$ for $j=i-2,i-1,i,i+1,i+2$.
We set $\tau=4.0,J=2.0,V\in\{10,15,20,25\}$, and tune $\alpha$ for the best performance. 
The first $K=10^4$ steps are used for training, and $400$ steps are predicted iteratively.
We employ the same parallel architecture for ESN-$n$, LSTM-$n$, and GRU-$n$ models, where $n \in \{80,100,120,150,500,1000\}$ is the number of nodes or hidden units in each group.
Figure~\ref{fig:ks:heat}(a) illustrates contours and error plots for the typical forecasts of HQR and ESN, which demonstrates the ability of the HQR to emulate the spatiotemporal chaos in approximately two Lyapunov time $\Lambda^{-1}_1$. 
Figure~\ref{fig:ks:heat}(b) illustrates the evolution of NRMSE averaged over 10 random tests for each model.
The HQR outperforms other models even with a small number of computational nodes.
This may be mainly due to the exponential numbers of degrees of freedoms behind the quantum measurements, which leads to the quantum computational supremacy region.

We want to note that our HQRC framework enables another flexible way to replicate the original chaotic system. 
In the conventional methodology, after initial training with data from the target system is done, one feeds the output from the readout part directly into the system's input module.  
However, it is equivalent to the modification of the linear feedback connection $\wfeed$ between the QRs.
Without loss of generality, we explain this statement by considering the simple case for one-dimensional output and the readout part without the bias term.
Let $\wout\in\cR_{1\times\ntot}$ denote the trained readout weights.
We replicate $\wout$ into $\wout^\prime \in \cR_{\nqrc\times\ntot}$ and represent the input in the closed system when the external input $u_k$ is withdrawn as
\begin{align}
    \bu^{\prime}_k = (1-\alpha)\wout^\prime\bz_{k-1} + \alpha\wfeed\bz_{k-1},
\end{align}
where $\bz_{k-1}\in\cR_{\ntot\times 1}$ is the reservoir states of HQR at $t=(k-1)\tau$. 
Therefore, if we set $\wfeed$ to 
$
\wfeed^\prime= (1-\alpha)\wout^\prime + \alpha\wfeed
$
after the training, the input becomes $\bu^{\prime}_k = \wfeed^\prime\bz_{k-1}$, and
the state evolution of the chaotic system will be replicated without requiring to feed the output from the readout part into each QR's input.

\subsection{Enhancing the noise robustness of chaos emulation via quantum-innate training}
\begin{figure}
	\centering
	\includegraphics[width=1.0\linewidth]{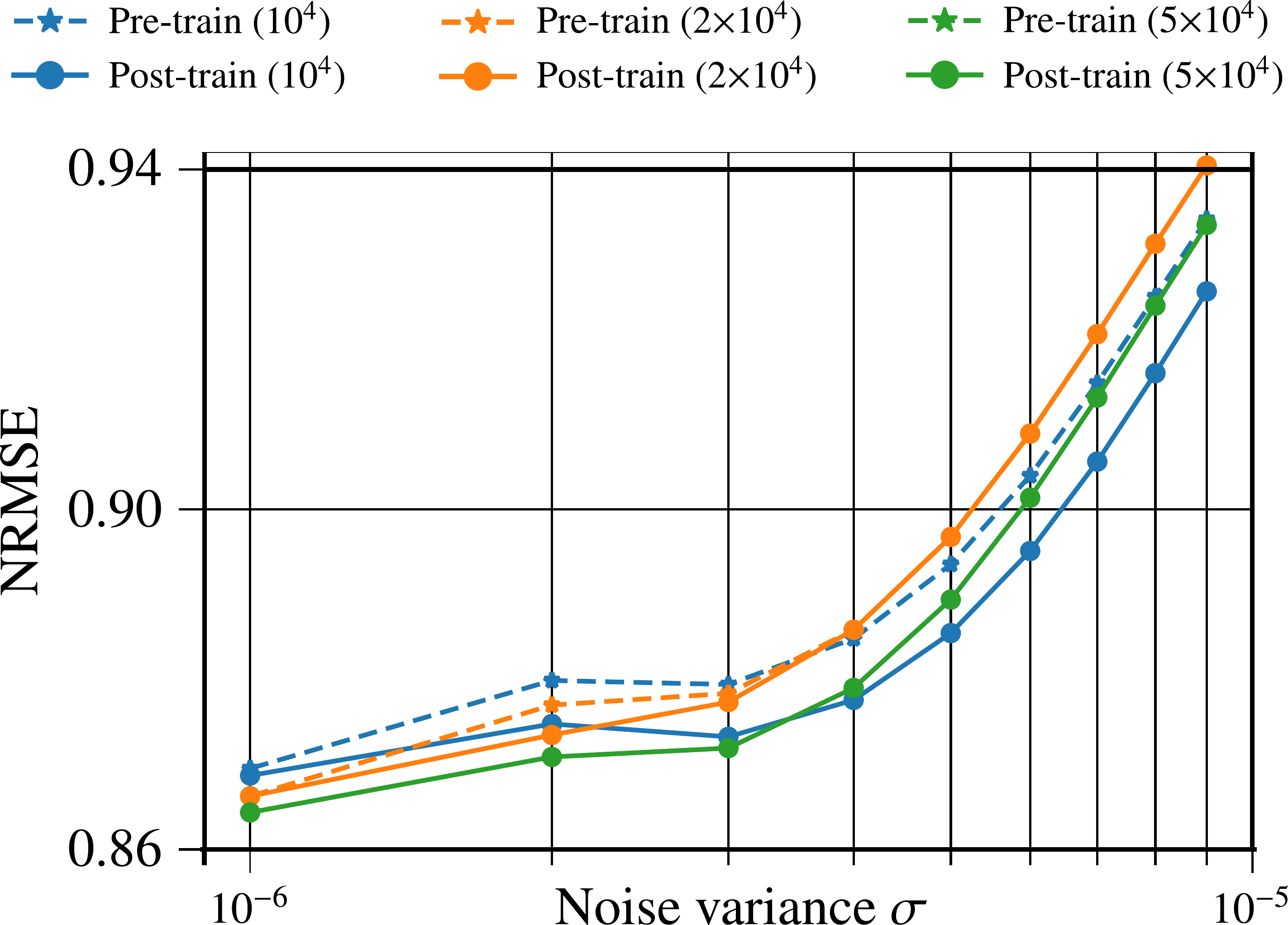}
	\caption{The NRMSE values averaged over 10 random prediction on the Lorenz system along different levels of noise added to the reservoir signals. The NRMSE values are evaluated in two regimes: without innate training (pre-train, dashed lines) and with innate training (post-train, solid lines). The number $K$ of time steps used in the main training are $10^4$ (blue), $2\times 10^4$ (orange), and $5\times 10^4$ (green).
	}\label{fig:innate:enhance:noise}
\end{figure}

In section~\ref{sec:innate:noise}, we introduce the quantum-innate training scheme based on FORCE learning. 
In this section, we show that the quantum-innate training scheme can enhance the noise robustness of the chaos emulation task.
For a simple demonstration, we consider the chaotic system of the Lorenz attractor mentioned in section~\ref{sec:chaos}.
Here, the first $K$ time steps are used for training, and $10^3$ time steps are predicted iteratively.
Our HQR-5 of six qubits, with $\tau=4.0, J=2.0$, and $V=1$, corresponds with $30$ computational nodes.
The connection strength $\alpha$ is set as $\alpha=0.0$ (without the feedback scheme) and $\alpha=0.1$.
We introduce the continuous Gaussian noise with a mean of zero and a variance of $\sigma$ into the reservoir signals at each time step.
In the training phase with data from the target chaotic system (the main training), this noise issue can be resolved via tuning the readout weights or adding some noise into the input. 
However, in the emulation phase, both noise signals from the feedback scheme and the closed-loop make the noise accumulating along with the system's evolution.
Therefore, if we introduce the quantum-innate training as the pre-train process to remove the reservoir signals' noise sensitivity, it will enable a more robust emulation of chaotic systems.

Figure~\ref{fig:innate:enhance:noise} demonstrates the average NRMSE over 10 random predictions with different $K$ time steps ($K=10^4$ (blue), $K=2\times 10^4$ (orange), and $K=5\times 10^4$ (green)) for the main training and different variance $\sigma$ of noise ($\sigma=10^{-6},\ldots,9\times 10^{-6}$).
The innate training is performed on the same training window with the main training of the target chaos.
The NRMSE values are evaluated in two regimes: without innate training (pre-train) and with innate training (post-train).
For each $K$ and $\sigma$, we plot the minimum NRMSE value in all settings of other parameters such as the connection strength $\alpha=0.0, 0.1$, the regularization parameter $\beta=10^{-7}, 10^{-9}$ in the Ridge regression of the main training (see ~\cite{gitsource:hqrc} for more details).
Figure~\ref{fig:innate:enhance:noise} shows that after 10 innate training loops (post-train) with learning rate $\zeta=10$,
the NRMSE values are reduced for all settings of $K$ and $\sigma$.
The reduction is significant when $K=10^4$ with all $\sigma \in [10^{-6}, 10^{-5})$ and when $K=10^5$ with $\sigma \in [10^{-6}, 3\times 10^{-6})$.
This result proves for the effectivity of the quantum-innate training scheme.
Increasing the number of main training steps will increase the system's noise robustness while the large noise variance leads to the problem of overfitting.

\section{Conclusion and discussion}
In this paper, we propose the HQRC framework, which offers an effective potential means for using quantum dynamics in machine learning tasks.
It is suggested that QRC has higher expressive power than classical RC, even for the same number of computational nodes; however, technical scalability is a major drawback.
Our framework introduces an implementation to solve this problem. Here, local operations are performed on each QR and the results of those operations are communicated in a classical manner.
The experimental results on emulating nonlinear systems, including high-dimensional spatiotemporal chaos, demonstrate that our framework can boost the computational power and scalability of QRC through both quantum and classical advantages.
Since only time evolution, according to the Hamiltonian as an interaction between nodes, is permitted in QRC, the design for an arbitrary nonlinear function remains as a future challenge.
A complete exploration of capabilities of other quantum systems as reservoirs and determining the optimal setting for the classical communications can be a possible direction for the future research.

The work most related to our proposal is Ref.~\cite{nakajima:2019:qrc}, which proposes the technique of spatial multiplexing, where multiple disjoint quantum systems can boost computational power of QRC. 
Although this approach is easy to implement in several conventional benchmark tasks, it does not increase the system's ability to represent complex dynamics.
Our HQRC is a general framework that enables a more flexible use of multiple quantum subsystems for reservoir processing.
In this way, it enables the introduction of some recent techniques to harness complex quantum dynamics by controlling the linear feedback weights, such as innate training~\cite{laje:2013:innate}, which provide a much broader range of applications.
We demonstrate in our paper that the quantum-innate training scheme can enhance the noise robustness when emulating the chaotic systems.
The more detailed investigation of the quantum-innate training's benefits in the current NMR ensemble system or other implementations remains an interesting research topic.

As a positive impact in the foreseeable future, our framework can be a typical case for quantum machine learning in the NISQ era for utilizing the noise-robust property of the classical system while exploiting the quantum computational supremacy region.
The flexibility of selecting the physical system as a computational resource, the low operation cost when running the hybrid quantum-classical system consisting of multiple small quantum systems, and exhibiting high computational power can enable our research to be utilized in real-world applications.
We believe that the experimental efforts in implementing physical quantum reservoir systems, such as the NMR ensemble system~\cite{negoro:2018:spins},
the random and noisy quantum circuits on super conducting quantum processors~\cite{chen:2020:temporal}, and the photonics system, can have the most benefit from our work.

\appendix
\section{Machine learning models for time series forecasting}

We briefly explain the conventional machine learning models used in the main text for time-series forecasting.
One can refer to Ref.~\cite{vlachas:2020:backpropagation} for more detailed explanations and implementations in the software framework.
Here, the models are trained on the time-series $\{\hbo_1, \ldots, \hbo_L\}$ of an observable $\hbo \in \cR_{d_o\times 1}$.
The temporal information of the observable is encoded via the internal high-dimensional state denoted by $\bh_t \in \cR_{d_h\times 1}$, where $d_h$ is the number of hidden units.
Given the time series $\{\hbo_1, \ldots, \hbo_t\}$, based on the hidden states, the models output the value $\bo_{t+1}$ as the prediction for the next target $\hat{\bo}_{t+1}$. This temporal processing can be modeled by the following equations:
\begin{align}
    \bh_t &= f_h^h(\hbo_t, \bh_{t-1}), \quad \hbo_{t+1} = f_h^o(\bh_t),
\end{align}
where $f_h^h$ is the hidden-to-hidden mapping and $f_h^o$ is the hidden-to-output mapping.

\subsection{Echo State Network (ESN)}
We consider the reservoir computing framework implemented in the context of echo state network (ESN)~\cite{jaeger:2004:rc}. Here, the hidden-to-hidden mapping $f_h^h$ is given by
\begin{align}
    \bh_t = \textup{tanh}(\bW_{h,i}\hbo_t + \bW_{h,h}\bh_{t-1}),
\end{align}
where $\bW_{h,i} \in \cR_{d_h\times d_o}$, and $\bW_{h,h} \in \cR_{d_h\times d_h}$ are fixed randomly.
The elements of $\bW_{h,i}$ are generated from a uniform distribution in $[-1.0, 1.0]$.
Moreover, to make the system satisfy the Echo State Property, the matrix $\bW_{h,h}$
is often set as a large low-degree matrix with its spectral radius (absolute value of the largest eigenvalue) is in a finite range ($< 1.0$). This condition can be satisfied by properly normalizing the elements in $\bW_{h, h}$.
The hidden-to-output mapping is set to
\begin{align}
    \bo_{t+1} = \bW_{o,h}\bh_t,
\end{align}
where $\bW_{o,h} \in \cR_{d_h\times d_o}$ is trained via regularized least-squared regression.
In practical applications, to enable the stability of ESN in long-term forecasting, Gaussian noise sampled from $\cN(0, \eta_n\sigma_d)$ is added to the training data. Here, $\sigma_d$ is the standard deviation of the data and $\eta_n$ is the noise level.
In our experiments for timeseries forecasting tasks, we set $\eta_n$ as a tuning parameter in $\{0, 10^{-3}, 5\times 10^{-3}\}$.

\subsection{Long Short-Term Memory (LSTM)}
The long short-term memory (LSTM)~\cite{hochreiter:1977:LSTM} was proposed to deal with the vanishing gradient problem of classical RNNs by utilizing the mechanism that allow information to be forgotten.
In LSTM, the hidden-to-hidden mapping is defined by the following recurrent functions:
\begin{alignat}{2}
    \bg_t^f &= \sigma_f(\bW_f\bq_t + \bb_f), \quad && \bg_t^i = \sigma_i(\bW_i\bq_t + \bb_i),\\
    \tilde{\bc}_t &= \textup{tanh}(\bW_c\bq_t + \bb_c), \quad && \bc_t = \bg_t^f \odot \bc_{t-1} + \bg_t^i \odot \tilde{\bc}_t,\\
    \bg_t^o &= \sigma_h(\bW_h\bq_t + \bb_h), \quad && \bh_t = \bg_t^o \odot \textup{tanh}(\bc_t),
\end{alignat}
where $\odot$ denotes the element-wise product, and $\bg_t^f, \bg_t^i, \bg_t^o \in \cR_{d_h\times 1}$ are the forget, input, and output gates signals, respectively. 
$\bq_t = [\bh_{t-1}, \hbo_t]\in \cR_{(d_h + d_o)\times 1}$ is the concatenated vector of the hidden state $\bh_{t-1}$ and the observable input $\hbo_t$, $\bc_t \in \cR_{d_h\times 1}$ is the cell state.
$\bW_f, \bW_i, \bW_c, \bW_h \in \cR_{d_h \times (d_h + d_o)}$ are weight matrices, and $\bb_f, \bb_i, \bb_c, \bb_h \in \cR_{d_h\times 1}$ are bias vectors.
The activation functions $\sigma_f, \sigma_i$, and $\sigma_h$ are sigmoid functions. Similar to ESN, the hidden-to-output mapping is set to
$
    \bo_{t+1} = \bW_{o,h}\bh_t,
$
where $\bW_{o,h} \in \cR_{d_h\times d_o}$.
In our study, we use back-propagation through time (BPTT) algorithm to train LSTM and its variations.
We refer Ref.~\cite{vlachas:2020:backpropagation} for a more detailed explanation of BPTT in the software framework.

\subsection{Gated Recurrent Unit (GRU)}
The gated recurrent unit (GRU)~\cite{choetal:2014:learning} is a variation of the LSTM without output gates.
GRU uses update gate and reset gate to decide what information should be passed to the output; thus, information from long ago can be kept in the training process without excluding information that is irrelevant to the prediction.
The recurrent mappings of the GRU are given by
\begin{alignat}{2}
    \bz_t &= \sigma_g(\bW_z\bq_t + \bb_z), \quad &&\br_t = \sigma_g(\bW_r\bq_t + \bb_r),\\
    \tilde{\bh}_t &= \textup{tanh}(\bW_h\bp_t + \bb_h), \quad &&\bh_t=(\boldsymbol{1}-\bz_t)\odot \bh_{t-1} + \bz_t \odot \tilde{\bh}_t,
\end{alignat}
where $\bz_t \in \cR_{d_h\times 1}$ is the update gate vector, $\br_t \in \cR_{d_h\times 1}$ is the reset gate vector, $\bq_t = [\bh_{t-1}, \hbo_t] \in \cR_{(d_h + d_o)\times 1}$, $\bp_t = [\br_t \odot \bh_{t-1}, \hbo_t] \in \cR_{(d_h + d_o)\times 1}$ are concatenated vectors.
The gating activation $\sigma_g$ is a sigmoid function, while $\bW_z, \bW_r, \bW_h \in \cR_{d_h \times (d_h + d_o)}$ are weight matrices and $\bb_z, \bb_r, \bb_h \in \cR_{d_h\times 1}$ are bias vectors.
The hidden-to-output mapping is set to
$
    \bo_{t+1} = \bW_{o,h}\bh_t,
$
where $\bW_{o,h} \in \cR_{d_h\times d_o}$.

\section{Acknowledgment}

We thank Y. Hasegawa, T. Van Vu, V. T. Vo, and T. Goto for insightful discussions and P. R. Vlachas for providing the instructions to run the framework in Ref.~\cite{vlachas:2020:backpropagation}.
We thank Y. Hasegawa and K. Inoue for providing the additional computational resources to perform our numerical experiments.

This work was based on results obtained from a project commissioned by the New Energy and Industrial Technology Development Organization (NEDO). Q. H. T and K. N. were supported by MEXT Quantum Leap Flagship Program (MEXT Q-LEAP) Grant No. JPMXS0120319794.

\bibliography{main.bib}

\end{document}